\documentclass[pra,twocolumn]{revtex4} 
\usepackage{graphicx}
\usepackage{amssymb, amsmath}
\usepackage{subfigure}

\begin{document} 
\date{\today}
\title{Superradiance of Harmonic Oscillators}
\author{M. Delanty}
\affiliation{Center for Engineered Quantum Systems, Department of Physics and Astronomy, Macquarie University, Sydney, NSW 2109, Australia}
\affiliation{CSIRO Materials Science and Engineering, P.O. Box 218, Lindfield, New South Wales 2070, Australia}
\author{S. Rebi\'c and J. Twamley}
\affiliation{Center for Engineered Quantum Systems, Department of Physics and Astronomy, Macquarie University, Sydney, NSW 2109, Australia}

\begin{abstract}

Superradiance, the enhanced collective emission of light from a coherent ensemble of quantum systems, has been typically studied in atomic ensembles. In this work we study the enhanced emission of energy from coherent ensembles of harmonic oscillators. We show that it should be possible to observe harmonic oscillator superradiance in a variety of physical platforms such as waveguide arrays in integrated photonics and resonator arrays in circuit QED. We find general conditions specifying when emission is superradiant and subradiant and find that superradiant, subradiant and dark states take the form of multimode squeezed coherent states and highly entangled multimode Fock states. The intensity, two-mode correlations and fraction of quanta trapped in the system after decay are calculated for a range of initial states including multimode Fock, squeezed, coherent and thermal states.  In order to explore these effects, the Law and Eberly protocol [C. K. Law and J. H. Eberly, Phys. Rev. Lett. \textbf{76}, 1055 (1996)] is generalized to prepare highly entangled multimode Fock states in circuit QED.
\end{abstract}

\maketitle

\section{Introduction}

A range of new and exciting quantum technologies have emerged recently that demonstrate quite detailed levels of coherent quantum control. In particular, exquisite quantum control has been achieved in highly versatile systems such as in circuit QED and integrated photonics. In integrated photonics these achievements include the creation of the first controlled-NOT (cNOT) quantum gate \cite{IPCnot} and the demonstration of quantum walks of correlated photons \cite{IPRandomWalks}. Furthermore, the Grover search and Deutsch-Jozsa quantum algorithms  \cite{Dicarlo} and the preparation of NOON, GHZ  and W states have recently been experimentally demonstrated in circuit QED \cite{MartinisNoon, GHZTransmon, GHZPhaseQubits}. Given these proof of principal experiments and improving fabrication techniques, next generation experiments have been looking towards many-body quantum systems \cite{GirvinSymmetryBreaking}. 

However, many-body quantum systems are exceptionally difficult to study experimentally. Issues surrounding initial state preparation, complex many-body evolution and the impossibility of full tomography have limited these experiments to all but the simplest of systems. From a theoretical perspective there are also difficulties in solving the complex equations decribing many-body system evolution. As most practical experimental systems are few-body quantum systems, $N < 20$, it is not possible to study interesting condensed matter phenomena such as the Kondo effect and quantum phase transitions in spin glasses. Given the technological ability, we seek to study new physics that can be explored in few-body experiments.

One such phenomena is harmonic oscillator superradiance. Superradiance is the enhanced emission from an ensemble of quantum systems due to field mediated build up of correlations within the ensemble. Initially studied in two level atomic ensembles by Dicke in 1954 \cite{Dicke}, superradiance has recently experienced a revival of interest with its observation  in quantum dots \cite{SR_QDs} and Bose Einstein condensates \cite{SRBEC2003}. However, superradiance has never been observed from continuous variable systems.

In this paper, we show that recent advances in superconducting quantum optics \cite{clarkecircuitQEDReview, GirvinReview} and integrated quantum photonics \cite{OBrienIPReview, ObrienNatureReview} provide an ideal environment for the practical exploration of harmonic oscillator superradiance phenomena. This paper is organized as follows; in section \ref{SRSectionNearIntro} superradiance of harmonic oscillators and its various physical implementations are discussed. We introduce a collective basis, analogous to the Dicke basis, and compare the initial emission intensities of atomic and oscillator ensembles in section \ref{system}. In section \ref{cohSec} a range of collective initial states are introduced and their preparation is considered in circuit QED and integrated photonics. The superradiance master equation is derived in section \ref{SRSection}, from which the emission intensity of an ensemble of oscillators is found. Decoherence free subspaces are also considered; we find a set of dark squeezed coherent states which do not decay and may be useful as a quantum memory for continuous variable quantum information. In section \ref{SRcriterionSection} we develop a general criterion to classify any state as superradiant, subradiant or normal radiant. The emission intensity, fraction of trapped energy and two-mode correlations are derived for a variety of initial states in section \ref{SRIc}. Using the superradiance criterion, superradiant parameter regimes are found for these initial states. Finally, specific physical implementations of harmonic oscillator superradiance in circuit QED and integrated photonics are considered in section \ref{imps}. In the circuit QED implementation, the Law and Eberly \cite{LawandEberly} protocol is generalized to prepare highly entangled multimode Fock states. This protocol is useful for tests of harmonic oscillator superradiance as it can be used to prepare many of the lower lying states of the collective basis.

\section{Superradiance \label{SRSectionNearIntro}}

\begin{figure}[t!]
\centering
\subfigure[]{
{\includegraphics[scale=0.23]{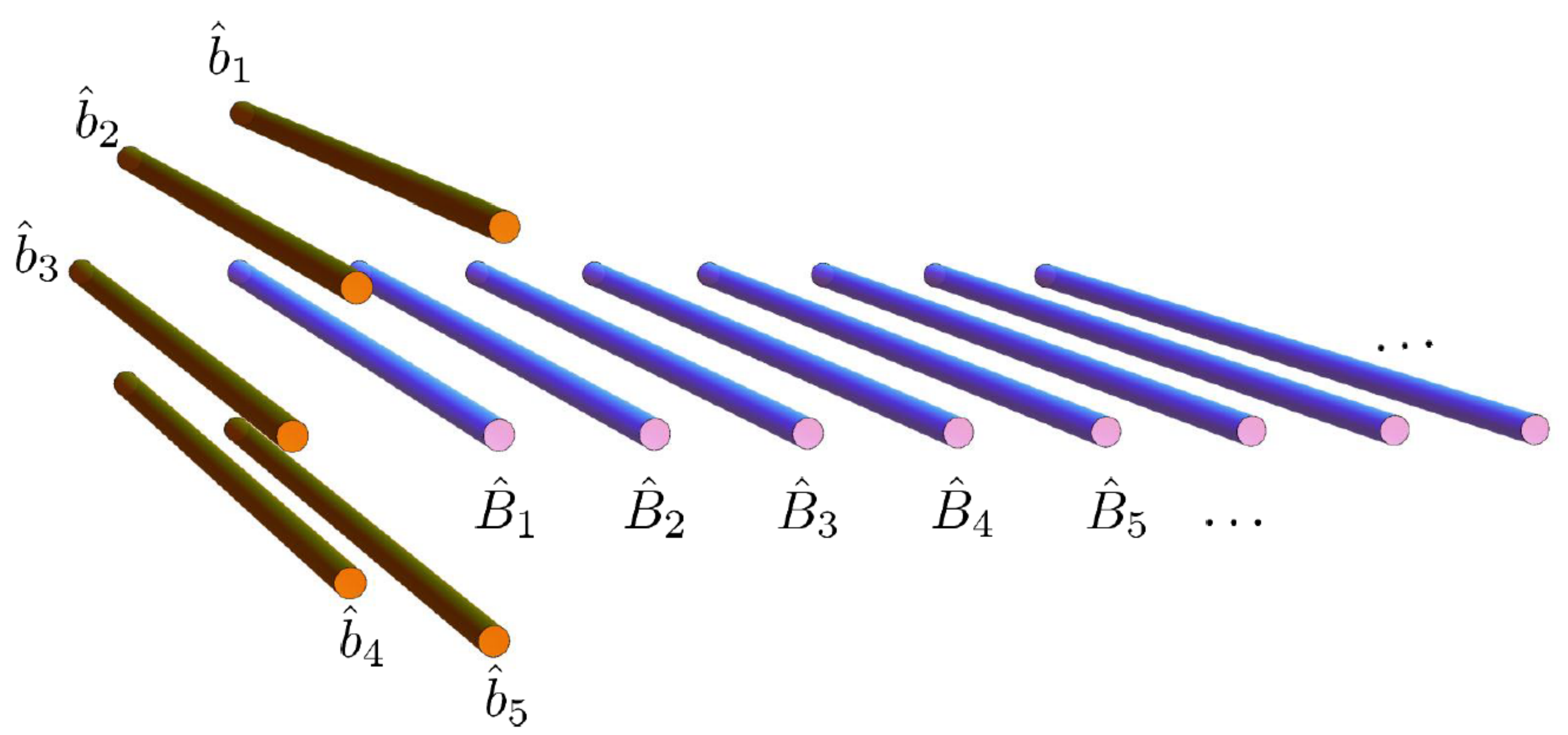} \label{IP_Diagram}}
}
\subfigure[]{
{\includegraphics[scale=0.25]{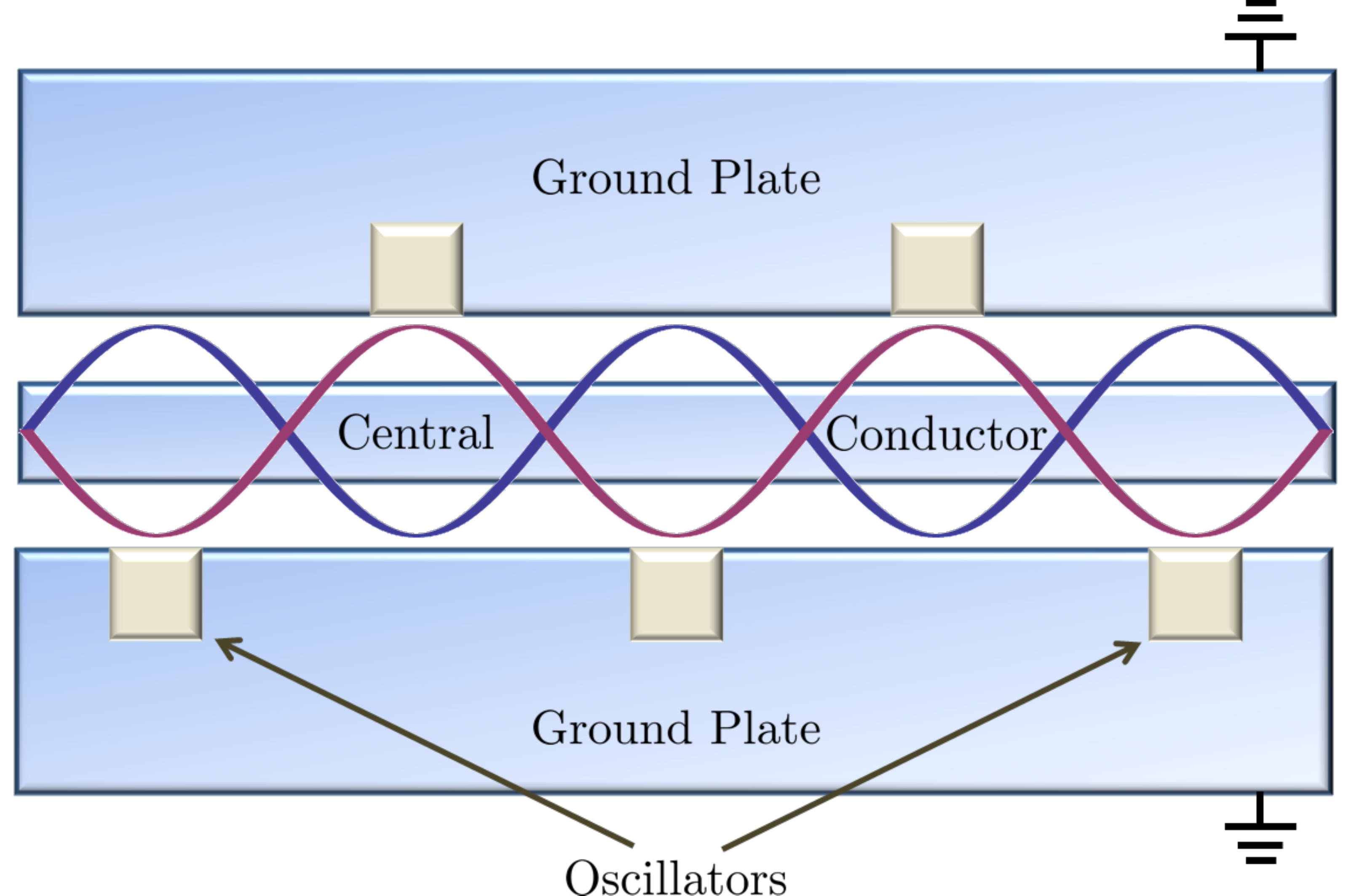}  \label{CircuitQEDDiagramSimple}}
}
\caption[]{ 
Diagrams of experimental implementations (a) \textit{Integrated Photonics}: a semicircular array of system waveguides, $\hat{b}_j$, collectively decay irreversibly into a semi-infinite linear array of bath waveguides, $\hat{B}_k$. By altering the input state of light into the system waveguides, $\hat{b}_j$, we can control the rate at which light couples into the semi-infinite array, $\hat{B}_k$. (b) \textit{Circuit QED}: an ensemble of oscillators (squares) coupled to a strongly damped stripline resonator. These oscillators could be the transmon qubit in the harmonic regime, superconducting stripline resonators or lumped element resonators.
}
\end{figure}

In this section we briefly review relevant experimental and theoretical work on superradiance and discuss several physical implementations where harmonic oscillator superradiance could be observed. Superradiance has been experimentally observed in a variety of systems including gases \cite{HarocheReview}, ions \cite{SR_Ions}, thin crystal slabs \cite{SRSlab}, quantum dots \cite{SR_QDs} and Bose Einstein condensates \cite{SRBEC2003}. Furthermore, numerous theoretical works have considered superradiance in a number of physical systems such as nitrogen vacancy centers in diamond and circuit QED \cite{MySRpaper, Agarwal1974,HarocheReview, BrandesReview}.

Experiments performed to date typically observe the classic $I \propto N^2$ intensity character of superradiance, however, complicated many-body effects usually inhibit a detailed study of the dynamics of emission or the build-up of quantum correlations. Experimental difficulties in typical superradiance studies arise from three sources: initial state preparation, intrinsic losses and measurement. Initial state preparation is problematic as superradiant Dicke quantum states are highly entangled many-body quantum states.  Most experiments cannot prepare such states and instead fire a pulse into the medium randomly exciting subsections of the ensemble \cite{HarocheReview,SRSlab}. Propagation effects through the ensemble can also lead to delays and distortions of the superradiant emission. Secondly, dephasing and dissipative losses can arise due to collisions within the ensemble or via interactions with lattice phonons which break the coherence required for superradiant decay. Finally, measurement is restricted to the observation of the emitted radiation rather than direct measurement of the ensemble dynamics, which makes a faithful reconstruction of the many-body dynamics difficult.

The vast majority of work on superradiance has focused on ensembles of two level systems. There has also been some work on superradiance of multi-level atoms  where more complex phenomena can be observed \cite{HarocheReview}. Such phenomena includes superradiant light beating, fluctuations between different light polarizations during superradiant emission and superradiant emission of successive pulses of different frequencies. Superradiance from an ensemble of harmonic oscillators was studied theoretically in the 1970's  \cite{PuriLawande, KatrielandAdam} most notably by Agarwal using phase space methods  \cite{AgarwalQHOpaper, AgarwalSRpaper,Agarwal1974, StojansPaper}. However, without a readily available experimental implementation, interest in superradiance of harmonic oscillators soon subsided \footnote{ However, we have recently become aware of \cite{ciracNJP}. }.

We now consider experimental systems where harmonic oscillator superradiance could be observed. Firstly we require an oscillator ensemble. Such ensembles could include waveguides in integrated photonics (Fig.~\ref{IP_Diagram}),  superconducting qubits in the harmonic regime \cite{TransmonPaper}, superconducting stripline \cite{GirvinReview} and lumped element resonators \cite{LumpedElement} (Fig.~\ref{CircuitQEDDiagramSimple}), mechanical resonators, photonic band gap cavities, silicon toroidal microresonators and micro discs, among others. Secondly we require the ensemble to decay into a common Markovian bath. This could be achieved in one of two ways; either all oscillators are located in a small region and interact with a common bath (Fig.~\ref{IP_Diagram}) or all oscillators interact with a common, strongly damped mode (Fig.~\ref{CircuitQEDDiagramSimple}). This requirement for a common bath is experimentally difficult and is the primary reason why harmonic oscillator superradiance has not been observed to date. Thirdly, measurement of the intensity from the ensemble is required, either directly by observing the emission or indirectly by measuring the energy loss from the ensemble.  Furthermore, desirable attributes of an experimental implementation would include the ability to prepare the oscillators in a variety of interesting states and the ability to measure two-mode correlations within the ensemble during decay. In this paper, it is shown that the above requirements are satisfied in an integrated photonics and circuit QED system.

\section{Bosonic Dicke Basis \label{system}}

\begin{figure}[t!]
\centering
\subfigure[]{
{\includegraphics[scale=0.34]{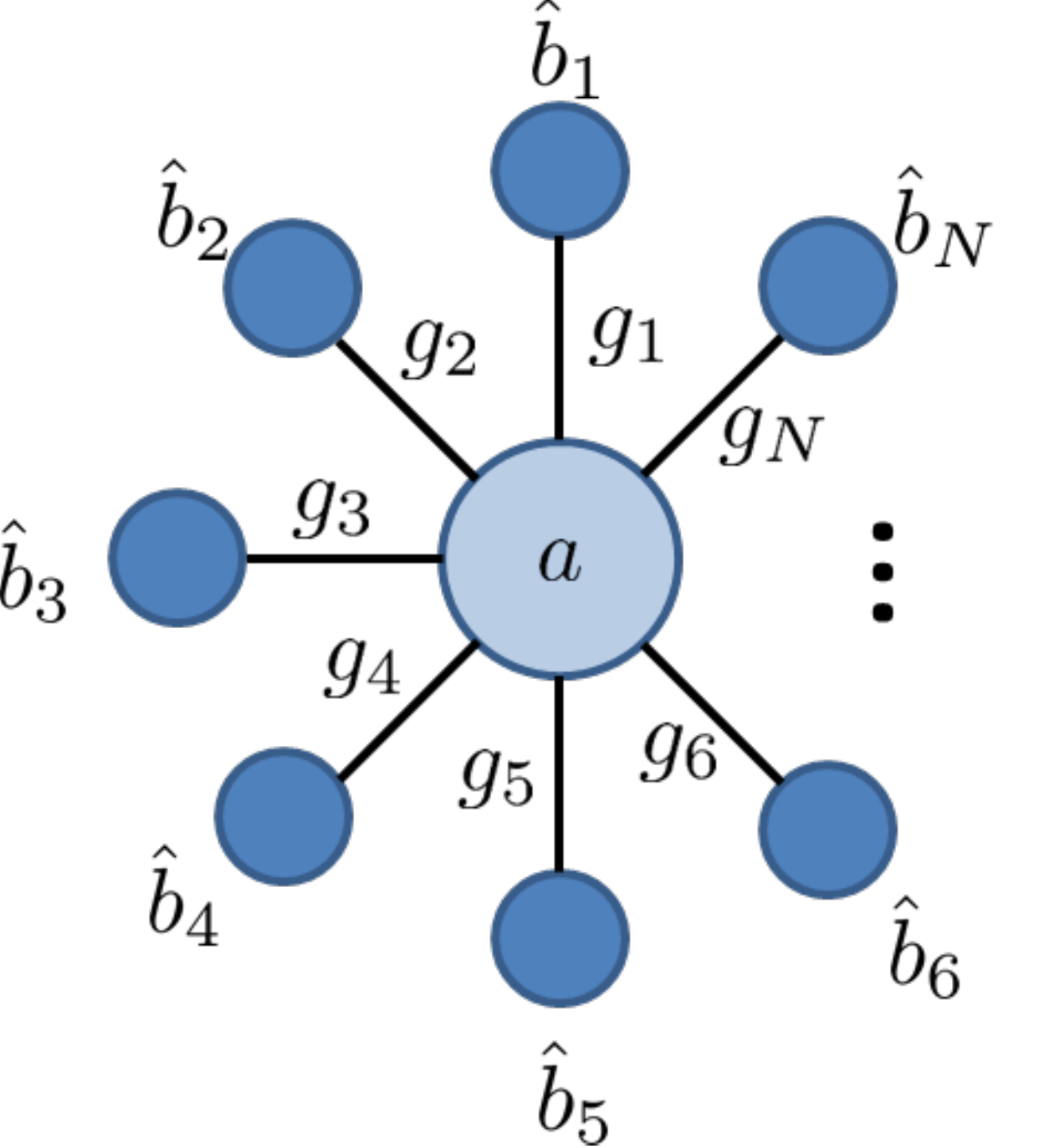} \label{StarSystem}}
}
\subfigure[]{
{\includegraphics[scale=0.25]{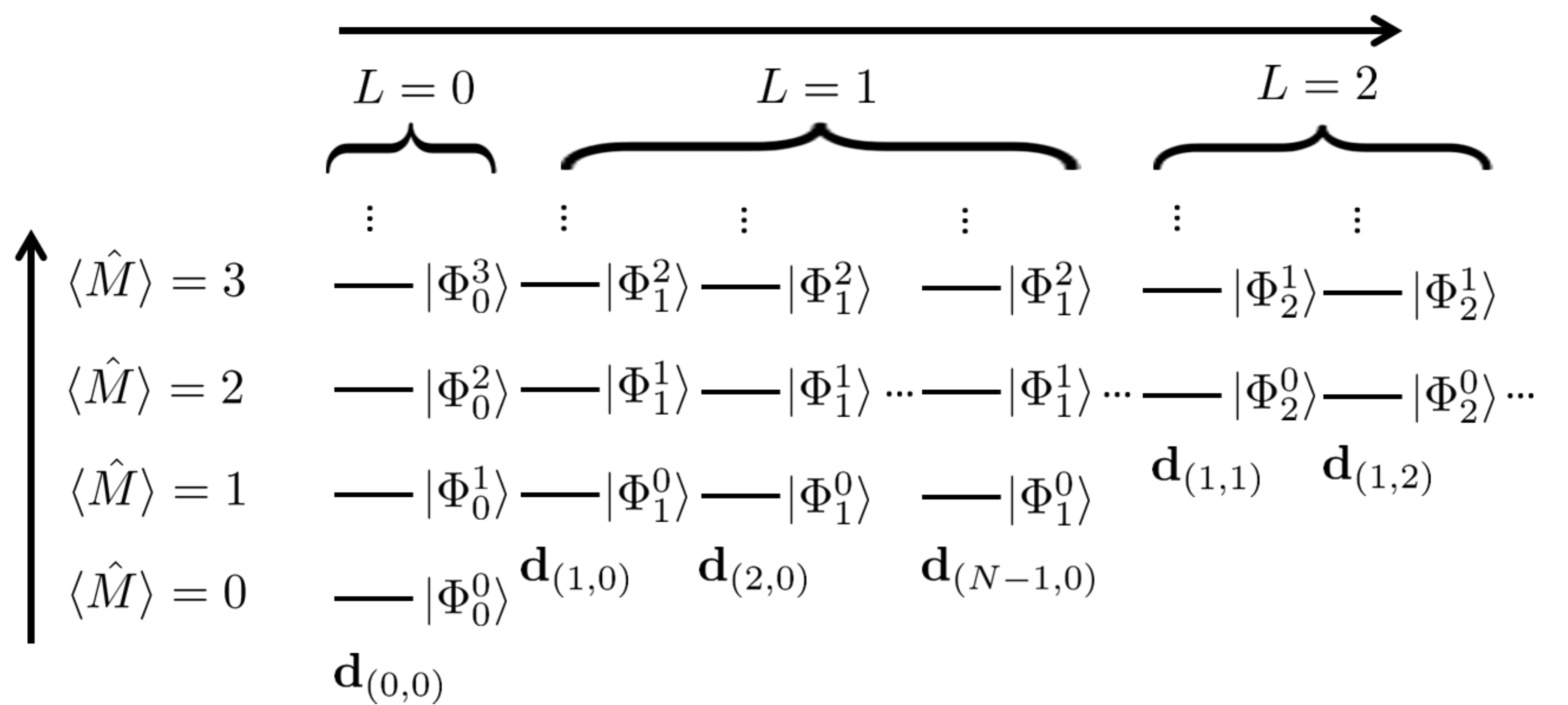} \label{EnergyLevels} }
}
\caption{ (a) Star shaped geometry in the Hamiltonian (\ref{Ham}). (b) Energy level diagram of the bosonic Dicke basis states, $| \mathbf{d}_{(i,j)}, \Phi^R_L \rangle$. As all states on each ladder have the same degeneracy vector, this vector is placed at the base of each ladder. The degeneracy vector at the bottom of each ladder is the sum of two unit vectors, $\mathbf{d}_{(i,j)}=\mathbf{e}_{i}+\mathbf{e}_{j}$, where we assume $\mathbf{e}_{0}$ is the zero vector. For example: $ | \mathbf{d}_{(2,0)}, \Phi^2_1 \rangle = (\hat{C}^{\dagger}_N)^2 \hat{C}^{\dagger}_2 | 0 \rangle^{\bigotimes N}/\sqrt{2}$ and $ | \mathbf{d}_{(1,2)}, \Phi^1_2 \rangle = \hat{C}^{\dagger}_N \hat{C}^{\dagger}_1 \hat{C}^{\dagger}_2 | 0 \rangle^{\bigotimes N}$.
}
\end{figure}

In this section we introduce the bosonic Dicke basis and compare atomic and oscillator superradiance. The bosonic Dicke basis arises in the study of an ensemble of oscillators coupled to a central mode (Fig.~\ref{StarSystem}). This is analogous to introducing the atomic Dicke basis in the study of the Tavis-Cummings model. The Hamiltonian for the star shaped oscillator system in Fig.~\ref{StarSystem} is,
\begin{equation} \label{Ham}
H/\hbar  = \omega \hat{a}^{\dagger} \hat{a} +   \sum^N_{j=1} \omega \hat{b}_j^{\dagger} \hat{b}_j +  \sum^N_{j=1} g_j(\hat{a}^{\dagger} \hat{b}_j + \hat{a} \hat{b}_j^{\dagger}  ),
\end{equation}
where, all oscillators have the angular frequency, $\omega$, and the $N$ oscillators couple to the central mode at the rates, $g_j>0$. Furthermore, the modes are assumed to be independent with $[\hat{a}, \hat{a}^{\dagger}] =1$, $[\hat{b}_i , \hat{b}_j^{\dagger} ]=\delta_{i,j}$ and all other commutators vanishing. This system can be easily realized in a variety of physical systems including the circuit QED system in Fig.~\ref{CircuitQEDDiagramSimple} \cite{KimMultisplitter}.  The Hamiltonian ($\ref{Ham}$) is diagonalized in Appendix A.

The Hamiltonian  can be rewritten using collective operators,
\begin{equation}\label{HGTc}
H/ \hbar = \omega \hat{a}^{\dagger} \hat{a} + \omega (\hat{L}+ \hat{R})+ \mathcal{G}_N (\hat{a}^{\dagger} \hat{C}_N + a \hat{C}_N^{\dagger}  ),
\end{equation}
where we have defined: the collective ladder operator $\hat{C}_N = \sum^N_{j=1} g_j \hat{b}_j/\mathcal{G}_N$, with $\mathcal{G}_j = \sqrt{\sum _{k=1}^j g_k^2}$; an operator to quantify the position on the ladder $\hat{R} = \hat{C}^{\dagger}_{N} \hat{C}_N$, an operator to label distinct ladders  $\hat{L} = \hat{M} - \hat{R}$ and an operator that quantifies the number of quanta in the $\hat{b}_j$ modes  $\hat{M} =  \sum^N_{j=1}  \hat{b}_j^{\dagger} \hat{b}_j$ (see Fig.~\ref{EnergyLevels}). These operators are analogous to Dicke's collective atomic operators \cite{carmichaelStatisticalMethods1},  $\hat{J}_i = \sum^{N}_{j=1} \hat{\sigma}^i_j$ and  $\hat{\mathbf{J}}^2 = \hat{J}^2_x +\hat{J}^2_y + \hat{J}^2_z$, where $i=\{ +,-,x,y,z\}$ and $\hat{\sigma}^i_j$ denote the individual qubit Pauli matrices, with $\hat{\mathbf{J}}^2 \sim \hat{L}, \hat{J}_z \sim \hat{M}, \hat{J}_+ \sim \hat{C}^{\dagger}_N$, $\hat{J}_- \sim \hat{C}_N $ and $\hat{J}_+ \hat{J}_- \sim \hat{R}$ \footnote[1]{Although these operators serve a similar role as the ladder label, total energy, raising and lowering operators and rate of emission, respectively, there are significant differences between the atomic and bosonic operator eigenvalues and commutation relations.}.

We now introduce a multimode collective basis, the \textit{bosonic Dicke basis}, which provides a convenient bosonic analogue to the the atomic Dicke basis. To this end we define a unitary transformation \footnote[2]{We note that there is no physical meaning behind this choice of the $\hat{C}_k$ operators ($k\neq N$). Any unitary transformation of the $\hat{b}_j$ modes that generates $N-1$ operators and the $\hat{C}_N$ operator is sufficient for the creation of a bosonic Dicke basis. 
} of the $N$ modes ($\hat{b}_j$),
\begin{equation}
\hat{C}_k = \frac{1}{\mathcal{G}_k \mathcal{G}_{k+1}} \left( - \mathcal{G}_k^2 \hat{b}_{k+1}+g_{k+1} \sum^k_{j=1} g_j \hat{b}_j   \right),
\end{equation}
where, $k=1,2, \dots, N-1$, in addition to the previously defined $\hat{C}_N$.  As the transformation is unitary, it can be shown that $[\hat{C}_j, \hat{C}^{\dagger}_{j'}  ] = \delta_{j, j'}$ ($j, j' \in \{1, 2, \dots, N\} $ ) with all other commutators vanishing \cite{KatrielandAdam}. Noting $\hat{M} =  \sum^N_{j=1}  \hat{C}_j^{\dagger} \hat{C}_j$ in terms of these collective operators, it can be shown that $[\hat{M}, \hat{R}]=[\hat{L}, \hat{R}]=0$. Therefore we can define the following orthonormal simultaneous eigenkets in terms of the eigenvalues $R$ and $L$,
\begin{eqnarray}
\hat{R} | \mathbf{d}_L, \Phi^R_L \rangle &=& R | \mathbf{d}_L, \Phi^R_L \rangle, \\ 
\hat{L} | \mathbf{d}_L , \Phi^R_L \rangle &=&  L |\mathbf{d}_L, \Phi^R_L \rangle,
\end{eqnarray}
where,
\begin{equation}\label{BosonicDickeBasis}
 |  \mathbf{d}_L, \Phi^R_L \rangle = \frac{(\hat{C}^{\dagger}_N)^R}{\sqrt{R!}} \prod^{N-1}_{k=1} \frac{ (\hat{C}^{\dagger}_k)^{m_k} }{\sqrt{m_k !} } | 0 \rangle^{\bigotimes N},
\end{equation}
$m_k:=  \langle\mathbf{d}_L, \Phi^R_L |  \hat{C}^{\dagger}_{k} \hat{C}_k|\mathbf{d}_L, \Phi^R_L \rangle \in \mathbb{Z}^{+}$ and $\sum^{N-1}_{k=1} m_k  = L$.
   
The eigenstate $ |  \mathbf{d}_L, \Phi^R_L \rangle $ is $\binom{L+N-2}{L}$-fold degenerate with respect to $\hat{R}, \hat{L}$ and $\hat{M}$; labeled by a degeneracy vector $\mathbf{d}_L = (m_1, m_1, \dots, m_{N-1}) $. For notational convenience we write the degeneracy vector in terms of unit vectors, $ \mathbf{e}_k$, i.e.~$\mathbf{d}_L = \sum^{N-1}_{k=1} m_k \mathbf{e}_k$ and denote the $(N-1)$ zero vector by $\mathbf{e}_0$. In addition in this new basis the ladder operators have the particularly simple form, %Furthermore,
\begin{eqnarray}\label{cn}
\hat{C}_N |\mathbf{d}_L, \Phi^R_L \rangle &=& \sqrt{R} | \mathbf{d}_L, \Phi^{R-1}_{L} \rangle,  \\
\hat{C}^{\dagger}_N |\mathbf{d}_L, \Phi^R_L \rangle &=& \sqrt{R+1} | \mathbf{d}_L, \Phi^{R+1}_{L} \rangle.
\end{eqnarray}
For clarity, the lower lying states of the bosonic Dicke basis are presented in Fig.~\ref{EnergyLevels}.  It is clear that $R$ is the number of states from the bottom of each ladder and $L$ is conserved on each ladder. The vertical axis corresponds to the total number of quanta in the state, $\langle \hat{M} \rangle$, and it can be seen that $\langle \hat{M} \rangle =L+R$.

For example, when there are two oscillators the collective operators are, 
\begin{eqnarray}
\hat{C}_1 &=& \frac{1}{\mathcal{G}_2} (g_2 \hat{b}_1 - g_1 \hat{b}_2),\nonumber\\
\hat{C}_2 &=& \frac{1}{\mathcal{G}_2} (g_1 \hat{b}_1 + g_2 \hat{b}_2),\nonumber \\
\hat{M} &=& b^{\dagger}_1 \hat{b}_1 + \hat{b}^{\dagger}_2 \hat{b}_2 = \hat{C}^{\dagger}_1 C_1 + \hat{C}^{\dagger}_2 \hat{C}_2, \nonumber \\
\hat{R} &=& \frac{1}{\mathcal{G}^2_2} \left( g^2_1 \hat{b}^{\dagger}_1 \hat{b}_1 + g^2_2 \hat{b}^{\dagger}_2 \hat{b}_2 + g_1 g_2 (\hat{b}^{\dagger}_1 \hat{b}_2+\hat{b}^{\dagger}_2 \hat{b}_1) \right),  \nonumber \\
\hat{L} &=& \left( 1- \frac{g^2_1}{\mathcal{G}^2_2}  \right) \hat{b}^{\dagger}_1 \hat{b}_1 + \left( 1- \frac{g^2_2}{\mathcal{G}^2_2}  \right) \hat{b}^{\dagger}_2 \hat{b}_2 -   \frac{g_1 g_2}{\mathcal{G}^2_2} (\hat{b}^{\dagger}_1 \hat{b}_2+\hat{b}^{\dagger}_2 \hat{b}_1), \nonumber
\end{eqnarray}
where, $\mathcal{G}_2=\sqrt{g^2_1 + g^2_2}$. Furthermore, the lower lying states of the bosonic Dicke basis (\ref{BosonicDickeBasis}) for $N=2$ oscillators are,
\begin{subequations}
\begin{eqnarray}
\! \! \! \! | \mathbf{e}_0 , \Phi^0_0 \rangle &=& | 0,0\rangle, \\
\! \! \! \! | \mathbf{e}_0 , \Phi^1_0 \rangle &=& \frac{1}{\mathcal{G}_2} ( g_1 | 1,0\rangle + g_2 | 0,1\rangle ),\\
\! \! \! \! | \mathbf{e}_0 , \Phi^2_0 \rangle &=& \frac{1}{\mathcal{G}^2_2} ( g^2_1 | 2,0\rangle + g^2_2 | 0,2\rangle + \sqrt{2}g_1 g_2 | 1,1\rangle ),\\
\! \! \! \! | \mathbf{e}_1 , \Phi^0_1 \rangle &=& \frac{1}{\mathcal{G}_2} ( g_2 | 1,0\rangle - g_1 | 0,1\rangle ),\\
\! \! \! \! | \mathbf{e}_1 , \Phi^1_1 \rangle &=& \frac{1}{\mathcal{G}^2_2} \{ \sqrt{2} g_1 g_2(| 2,0\rangle - | 0,2\rangle) \nonumber \\ 
\! \! \! \!&+& (g^2_2 - g^2_1) | 1,1\rangle \}, \\
\! \! \! \! |2 \mathbf{e}_1 , \Phi^0_2 \rangle &=& \frac{1}{\mathcal{G}^2_2} ( g^2_1 | 0,2\rangle + g^2_2 | 2,0\rangle - \sqrt{2}g_1 g_2 | 1,1\rangle ).
\end{eqnarray}
\label{BosonicDickeNeq2}
\end{subequations}
In Fig.~\ref{EnergyLevels} these states are respectively, the bottom, first and second excited states on the left ladder; the bottom  and first excited state on the second ladder from the left; and the bottom state on the right ladder. Furthermore, for equal coupling rates, $g_1=g_2$, the left ladder states ($L=0$) can be expressed as,
\begin{eqnarray}
| \mathbf{e}_0 , \Phi^R_0 \rangle &=&  \frac{1}{\sqrt{2^R}} \sum^R_{n=0} \sqrt{ \binom {R} {n} } |R-n, n \rangle.
\end{eqnarray}

Given the above it is possible to compare the atomic and bosonic Dicke bases to understand the differences between atomic and oscillator superradiance. The bosonic Dicke states are simultaneous eigenstates of $\hat{R}$ and $\hat{L}$, whereas the atomic Dicke states are simultaneous eigenstates of the collective operators,  $ \hat{\mathbf{J}}^2 |l, m\rangle = l(l+1) | l, m\rangle$ and $\hat{J}_z | l, m\rangle = 2m |  l, m\rangle$, where, $N/2\ge l \ge |m| \ge 0$ \cite{carmichaelStatisticalMethods1}. Noting $\hat{M}=\hat{L}+\hat{R}$, we could have just as easily defined the bosonic Dicke basis as simultaneous eigenstates of $\hat{L}$ and $\hat{M}$ as the operator correspondences would imply, $\mathbf{J}^2 \sim \hat{L}, J_z \sim \hat{M}$. However, as we show in section \ref{SRSection}, $\hat{R}$ is significantly more important in harmonic oscillator superradiance.

\begin{figure}[t!]
\centering
\subfigure[]{
{\includegraphics[scale=0.26]{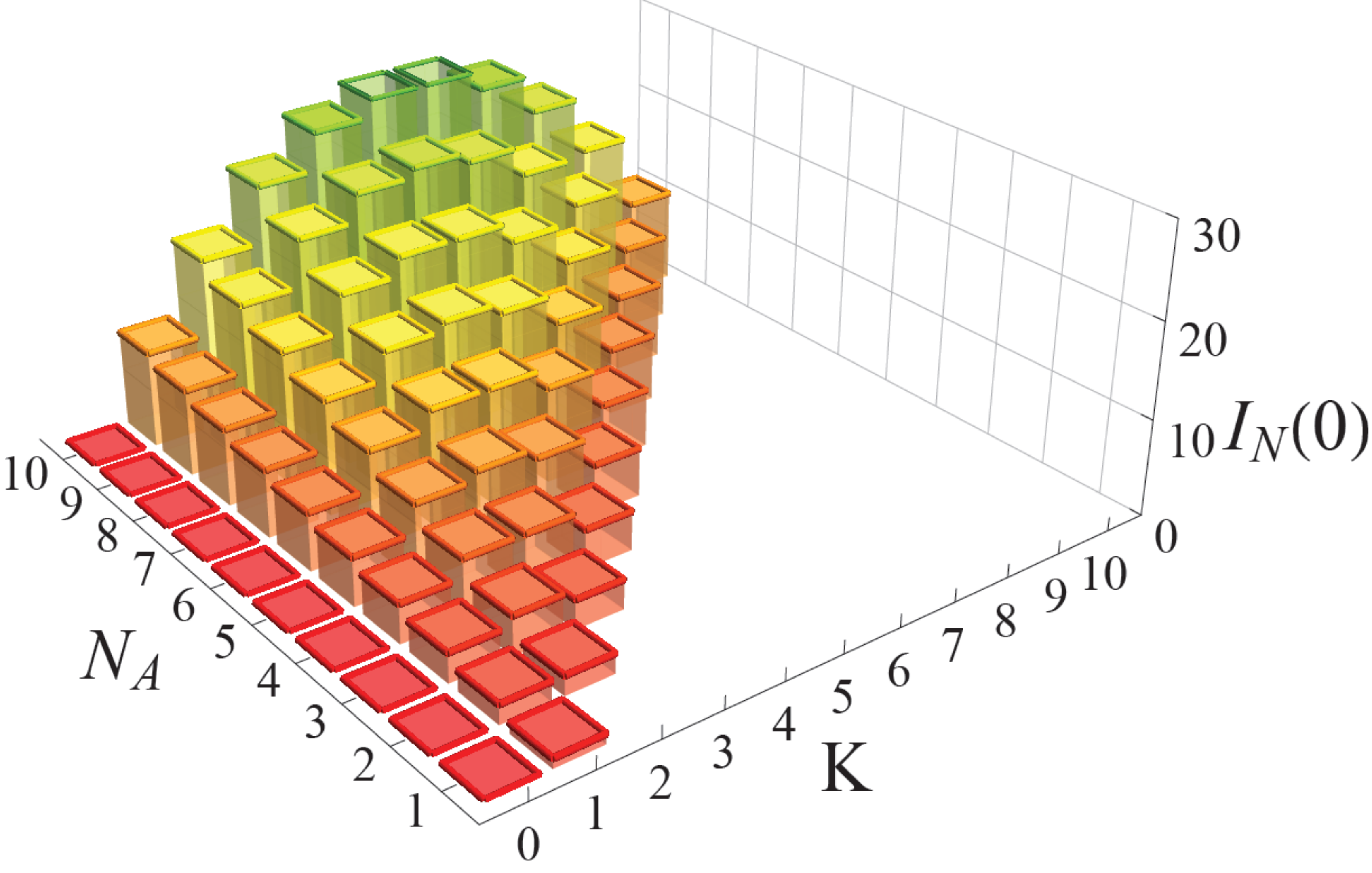}}
}
\subfigure[]{
{\includegraphics[scale=0.26]{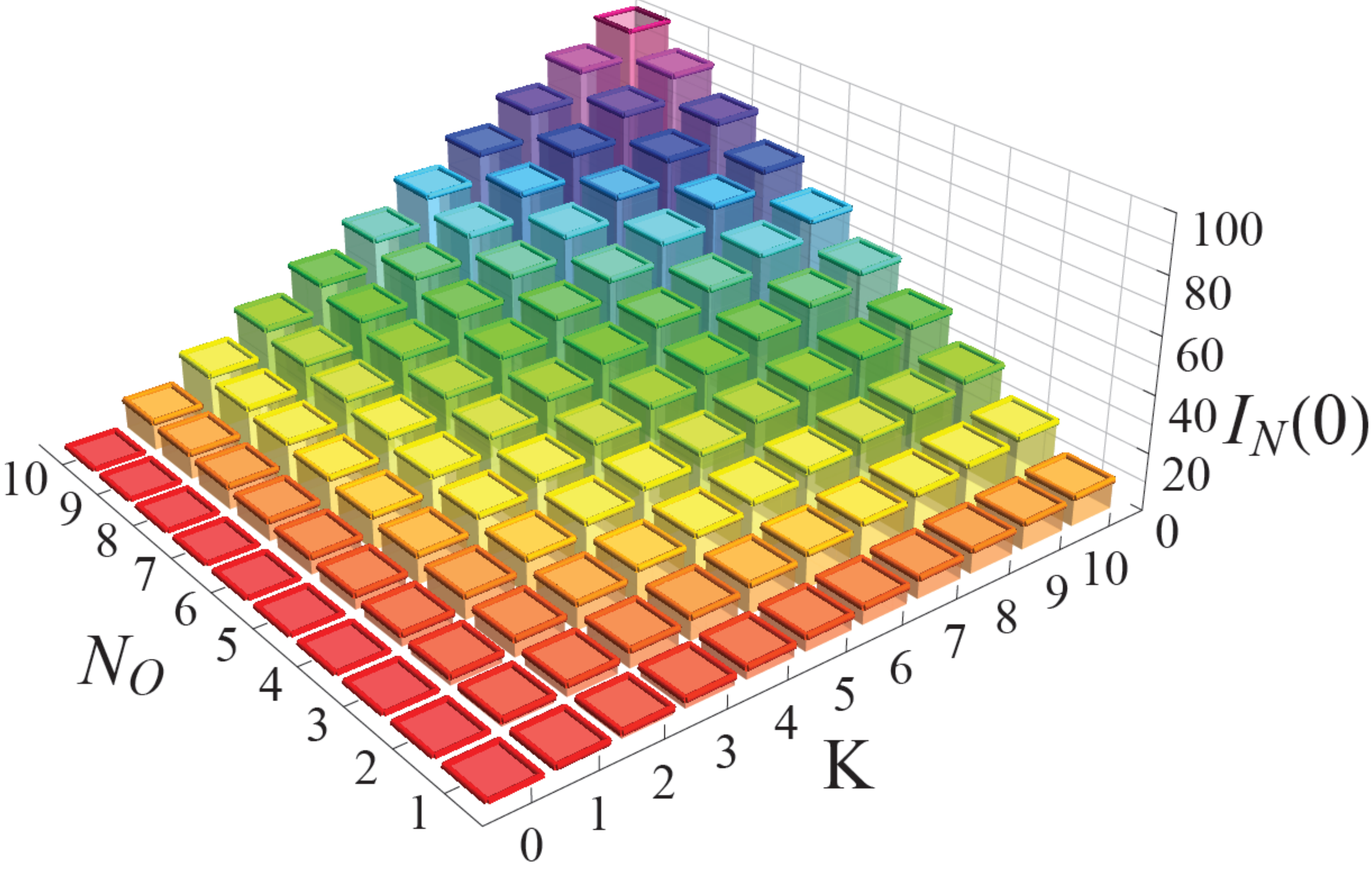}}
}
\caption[]{ Comparison of the maximum possible initial intensity, $I_N (0)$,  from an ensemble of (a) $N_A$ atoms and (b) $N_O$ oscillators with $K$ quanta in each initial state. The atoms are prepared in $|N/2,K-N/2\rangle$ ($N\ge K\ge 0$) and the oscillators are prepared in $| \mathbf{e}_0,\Phi^K_0 \rangle$ ($K\ge 0$). Note the substantial difference in scale in the vertical (intensity) axis as the oscillator emission is  significantly larger for large $K$. 
\label{Intensities}
 }
\end{figure}

Continuing the comparison of atoms and oscillators we note that the atomic Dicke basis has $2^N$ elements, while the bosonic Dicke basis has an infinite Hilbert space due to the infinite Hilbert space of each of the oscillators. In the atomic case there is a maximum excitation level for the atoms, $|N/2 , N/2 \rangle = |e \rangle^{\bigotimes N}$ and a finite number of quanta that can possibly be emitted from a finite ensemble of atoms, which does not exist in the bosonic case. This leads to large differences in the intensity of emission from either ensemble. In particular, the initial emission intensity for an ensemble of atoms prepared in the atomic Dicke state, $|l, m \rangle$ is \cite{Dicke}, 
\begin{eqnarray}
I_N (0) &\propto& \sum^N_{i,j=1} \langle l, m | \hat{\sigma}^+_i  \hat{\sigma}^-_j |l, m \rangle, \nonumber \\
&=& \langle l, m | \hat{J}_+ \hat{J}_- |l, m \rangle, \nonumber \\
&=& (l+m)(l-m +1). \label{IAtom}
\end{eqnarray} 
For $N$ atoms the intensity (\ref{IAtom}) is largest for the state $|N/2,0 \rangle$, where the initial intensity is $I_N (0) \propto \frac{N}{2}(\frac{N}{2}+1)$. This is the origin of the $I_N (0) \propto N^2$ hallmark of superradiance initially discovered by Dicke \cite{Dicke}. In contrast, the initial emission intensity for an ensemble of $N$ oscillators with identical coupling rates, $g_j = g \, \, \forall j$, prepared in the bosonic Dicke state, $|\mathbf{d}_L, \Phi^R_L \rangle$ is \cite{KatrielandAdam}, 
\begin{eqnarray}
I_N (0) &\propto& \sum^N_{i,j=1} \langle \mathbf{d}_L, \Phi^R_L | \hat{b}^{\dagger}_i \hat{b}_j |\mathbf{d}_L, \Phi^R_L \rangle, \nonumber \\
&=& N \langle \mathbf{d}_L, \Phi^R_L | \hat{C}^{\dagger}_N \hat{C}_N |\mathbf{d}_L, \Phi^R_L \rangle, \nonumber \\
&=& N R. \label{IFields}
\end{eqnarray} 
As $R$ is not bounded above like the atomic variables, $N/2\ge l \ge |m| \ge 0$, there is no fundamental upper limit to the emission intensity.

Another example that illustrates the differences between superradiance of atoms and oscillators is the comparison of the maximum initial intensity from an ensemble of $N$ atoms or oscillators with $K$ quanta in the initial state with $N\ge K$. For oscillators with identical coupling rates the state with the greatest initial emission intensity is $|\mathbf{e}_0, \Phi^K_0 \rangle $ with $I_N (0) \propto NK$. For $N$ atoms with $K$ excitations within the ensemble the greatest initial intensity occurs from the state $|N/2, K-N/2 \rangle$, where from (\ref{IAtom}) $I_N (0) \propto NK +K - K^2$. For $K=1$ the intensities coincide, however for $K>1$ the oscillators always have a greater initial intensity. Interestingly, in the large $N$ limit the intensities coincide.  The initial intensities from  $|N/2, K-N/2 \rangle$ and $|\mathbf{e}_0, \Phi^K_0 \rangle $ are plotted in Fig.~\ref{Intensities} for comparison. It is clear that the initial emission intensity for the system of harmonic oscillators is greater than or equal to the initial intensity from an equivalent ensemble of atoms. Furthermore the atoms have both a maximum possible intensity and a finite Hilbert space as there are only $(N+1)$ initial states.

\section{Collective Squeezed Coherent States and State Preparation in Integrated Photonics and Circuit QED  \label{cohSec}}

In this section we define continuous variable states in the bosonic Dicke basis and discuss how these states could be prepared in circuit QED and integrated photonics. As will be shown in section \ref{SRSection} these collective squeezed coherent states arise in the study of harmonic oscillator superradiance from the master equation's dependence on the $N$-th collective operator, $\hat{C}_N$. These states may also have applications in  multimode continuous variable quantum information \cite{CVQIReview}.

\subsection{Collective Squeezed Coherent States}

Here we introduce collective displacement and squeeze operators and expand the resulting squeezed coherent states in the bosonic Dicke basis. Collective coherent states can be generated via combinations of the collective displacement operators,  
\begin{eqnarray}
\hat{D}_{N} (\alpha) &=& \exp (\alpha \hat{C}_N^{\dagger} - \alpha^* \hat{C}_N)  
= \prod^N_{j=1} \hat{D}^s_j ( \frac{g_j}{\mathcal{G}_N} \alpha), \\
\hat{D}_{k} (\beta) &=& \exp (\beta \hat{C}_k^{\dagger} - \beta^* \hat{C}_k)  \\
&=& \hat{D}^s_{k+1} (-\frac{\mathcal{G}_{k}}{\mathcal{G}_{k+1}} \beta) \prod^k_{j=1} \hat{D}^s_j ( \frac{g_j g_{k+1}}{\mathcal{G}_k \mathcal{G}_{k+1}} \beta) ,\nonumber
\end{eqnarray}
where, $k\neq N$, and $\hat{D}^s_j (\alpha) =  \exp (\alpha \hat{b}_j^{\dagger} - \alpha^* \hat{b}_j)$ is the single mode displacement operator. Collective coherent states can be expanded in the bosonic Dicke basis, for example,
\begin{eqnarray}
\! \! \! \! \! \! \! \! \! \! \! \! | \Phi^{\alpha}_L \rangle &=&\hat{D}_{N} (\alpha)  |\mathbf{d}_L, \Phi^{0}_L \rangle =  e^{- \frac{|\alpha|^2}{2}} \sum^{\infty}_{R=0}  \frac{\alpha^R}{\sqrt{R!}}  |\mathbf{e}_0, \Phi^{R}_L \rangle, \label{coh1}  \\ 
\! \! \! \! \! \! \! \! \! \! \! \! |  \Phi^R_{\beta_k} \rangle  &=&\hat{D}_{k} (\beta)  | \mathbf{e}_0, \Phi^R_{0} \rangle =  e^{- \frac{|\beta|^2}{2}} \sum^{\infty}_{L=0}  \frac{\beta^L}{\sqrt{L!}}  | L \mathbf{e}_k, \Phi^R_{L} \rangle, \label{coh2} \\ %\label{Dk}  \\
 \! \! \! \! \! \! \! \! \! \! \! \!|  \Phi^{\alpha}_{\beta_k} \rangle &=&\hat{D}_{k} (\beta)  \hat{D}_{N} (\alpha)  |  \mathbf{e}_0, \Phi^0_{0}  \rangle   \nonumber \\
\! \! \! \! \! \! \! \! \! \! \! \! &=& e^{- \frac{|\alpha|^2+|\beta|^2}{2}} \sum^{\infty}_{R,L=0}  \frac{\alpha^R}{\sqrt{R!}}  \frac{\beta^L}{\sqrt{L!}}  | L \mathbf{e}_k, \Phi^{R}_L\rangle, \label{coh3}
\end{eqnarray} %the elements of $\mathbf{d}^k_L$ are $[\mathbf{d}^k_L]_n = L \delta_{k,n}$ and 
where the following identity was used, 
\begin{equation}
\hat{C}_k^{\dagger} | \mathbf{d}_L, \Phi^R_L\rangle = \sqrt{m_k +1}|\mathbf{d}_{L} + \mathbf{e}_k, \Phi^R_{L+1}\rangle \; \; \; \; (k\neq N),
\end{equation}
which can be easily verified using (\ref{BosonicDickeBasis}).
Displaced number states could also be generated with the collective displacement operators \cite{displacednumberstate}.

Experimentally, the collective coherent states that are the result of displacements of the  multimode vacuum, $|\mathbf{e}_0, \Phi^0_0 \rangle= |0\rangle^{\bigotimes N}$, can be prepared via a product of single mode coherent states, $| \alpha_1\rangle_1 | \alpha_2\rangle_2 \dots | \alpha_N\rangle_N$. For example, 
\begin{eqnarray}
|  \Phi^{\alpha}_{\beta_1} \rangle &=& \hat{D}_{N} (\alpha) \hat{D}_{1} (\beta) | 0 \rangle^{\bigotimes N}  \nonumber \\
&=& e^{i \phi} | \frac{g_1}{\mathcal{G}_N} \alpha  + \frac{g_2}{\mathcal{G}_2} \beta  \rangle_1 | \frac{g_2}{\mathcal{G}_N} \alpha  - \frac{g_1}{\mathcal{G}_2} \beta \rangle_2 \bigotimes^{N}_{j=3} | \frac{g_j}{\mathcal{G}_N} \alpha  \rangle_j \nonumber 
\end{eqnarray}
where, $\phi$ is an overall phase. Therefore we require the ability to prepare each mode independently with a well defined phase and amplitude relationship between the modes.

Although difficult to engineer, for the sake of completeness we now consider collective mode squeezed states. These can be generated via the single- and two-collective mode squeeze operators, 
\begin{eqnarray}
 \hat{S}_{j} (\xi ) &=&\exp \left(  \frac{\xi^*}{2} \hat{C}^2_j  - \frac{\xi}{2} (\hat{C}_j^{\dagger})^2 \right), \label{collSqOps} \\
 \hat{S}_{j,l} (\xi ) &=&\exp\left(   \xi^* \hat{C}_j \hat{C}_l  - \xi \hat{C}_j^{\dagger} \hat{C}_l^{\dagger} \right),
 \end{eqnarray}
where, $\xi = r  e^{i \theta}$. With these operators it is possible to generate a variety of squeezed coherent states and squeezed number states \cite{squeezedreview}. For example the $N$-th single collective mode squeezed vacuum is,
\begin{eqnarray} \label{SqucollVac}
| \Phi^{\xi}_0 \rangle &=& \hat{S}_{N} (\xi ) | \mathbf{e}_0 ,\Phi^0_0 \rangle  \nonumber \\ 
 &=& \frac{1}{\sqrt{\cosh r}} \sum^{\infty}_{R=0} \lambda_R e^{i R \theta} |\mathbf{e}_0 , \Phi^{2R}_0 \rangle,
\end{eqnarray}
where, $\lambda_R=(-1)^R \frac{\sqrt{(2R!)}}{2^R R!}  (\tanh r)^R$.

\subsection{State Preparation in Integrated Photonics}

We now consider how to prepare various initial states in the integrated photonics architecture. Firstly, the product coherent state $| \alpha_1, \alpha_2, \dots , \alpha_N\rangle$ can be prepared simply by driving the separate system waveguides with a specific amplitude and phase relationship. Injecting single mode squeezed light into each system mode is also feasible in current experiments \cite{PhotonicsNOONCohSq}. Similarly, the product Fock state, $| n_1, n_2, \dots n_N\rangle$, can be prepared by injecting Fock states from an external optical circuit. Furthermore, a recent experiment has demonstrated that single mode arbitrary states of light up to the two-photon level can be prepared \cite{TravellingArb2photonStates}. Using this protocol it is possible to create the unentangled state,
\begin{eqnarray}
|\psi \rangle = \bigotimes^{N}_{j=1} \sum^2_{n=0} c^{(j)}_n \frac{ (\hat{b}^{\dagger}_j)^{n}}{\sqrt{n!}} |0 \rangle_j .
\end{eqnarray} 

There is considerable difficulty in preparing the highly entangled multimode Fock states $|\mathbf{d}_L, \Phi^R_L\rangle$ and the collective squeezed coherent states. However there have been promising results. Recent progress in entangled multimode optical states have achieved the NOON state,  $(|N,0 \rangle + |0, N \rangle)/\sqrt{2}$, for $N=5$ \cite{PhotonicsNOONCohSq}. Furthermore, there is a proposal for generating multimode entangled states in waveguide arrays including a four path-two photon entangled state \cite{PathEntangledNOONTheory}.

\subsection{State Preparation in Circuit QED}

We now consider state preparation in the circuit QED architecture. Similar to integrated photonics the product coherent state $| \alpha_1, \alpha_2, \dots , \alpha_N\rangle$ can be prepared by driving the modes. There are also experimental methods of preparing single resonator Fock states \cite{MartinisLandE}, which could be used in parallel to create the product Fock state, $| n_1, n_2, \dots n_N\rangle$. Furthermore, there have been several proposals to generate entangled single mode and  two mode states in circuit QED. High fidelity arbitrary single mode Fock states, 
\begin{equation} \label{LandEstate}
| \psi \rangle = \sum^{N_1}_{n=0} c_n | n \rangle,
\end{equation}
have been experimentally demonstrated for up to $N_1=10$ using the Law and Eberly protocol \cite{LawandEberly} in a circuit QED architecture with a phase qubit and a coplanar waveguide resonator \cite{MartinisLandE}. The maximum Fock state that can be prepared, $N_1$, is limited primarily by decoherence. Improved qubit and resonator design and fabrication will allow larger Fock states to be prepared \cite{MartinisFock, MartinisLandE}. Preparation of continuous single mode states such as cat states and squeezed states have also been the subject of research \cite{CatCircuitQED,SqueezedCircuitQED}. 

There have also been proposals for two-mode states in circuit QED, however, these remain difficult to engineer. %and is state of the art/forefront of technology.
The focus of the research to date has been on NOON states, due to their application in metrology \cite{MartinisNoon,  PRL2Mode}. Recently, NOON states have been prepared experimentally for $N=1,2,3$ at a fidelity of $0.76, 0.50, 0.33$, respectively. NOON states with $N>3$ should be attainable with improved qubit coherence times \cite{MartinisNoon, Shoelkopf3D}. In addition, MOON states \footnote[3]{Using (\ref{FNSRcond}) it can be shown that MOON states are normal radiance. }, $(|M,0 \rangle + |0,N \rangle )/\sqrt{2}$, can be prepared with a similar protocol \cite{MartinisNoon}. The Law and Eberly protocol was recently generalized \cite{PRL2Mode,LawandEberly2} to prepare arbitrary two mode states, 
\begin{equation}\label{LandEstate2mode}
| \psi  \rangle = \sum^{N_1}_{n=0} \sum^{N_2}_{m=0} c_{n,m} | n, m \rangle,
\end{equation}
however, the fidelity of this state was not studied \footnote[4]{From the single mode experiment one expects $N_1, M_1 \ll 10$.}. This protocol is particularly important for tests of harmonic oscillator superradiance as it can be used to prepare many of the lower lying states in the bosonic Dicke basis (see for example (\ref{BosonicDickeNeq2})). Research into three mode entanglement is still in its infancy, however, there are promising results such as a recent experiment demonstrating coherent control of photons between three resonators \cite{Photonshellgame}. High fidelity entanglement of $N>2$ modes in circuit QED is still an open problem.

\section{Superradiance Master Equation and Decoherence Free Subspaces \label{SRSection}}

In this section we consider damping of the oscillators in Fig.~\ref{StarSystem} in order to provide a common bath for superradiant emission. It is assumed that each oscillator interacts with a zero temperature heat bath. We first state the master equation of the system. We next show that it is possible to prepare states of the oscillator ensemble that are decoherence free subspaces with respect to the central mode. Finally we derive the superradiance master equation and solve for the intensity of emission of the oscillator ensemble into the strongly damped central oscillator.

Assuming that each oscillator in Fig.~\ref{StarSystem} interacts with a zero temperature Markovian heat bath leads to non-unitary evolution of the system. The system dynamics are now described by the following master equation, 
\begin{equation}\label{MEGTcB4gammaeq0}
\dot{\rho} = -\frac{i}{\hbar} [H, \rho] + \frac{\kappa}{2}\mathcal{D}[\hat{a}]\rho  +   \sum^N_{j=1}  \frac{\gamma_j}{2} \mathcal{D}[\hat{b}_j]\rho, 
\end{equation}
where, the Hamiltonian is defined in (\ref{HGTc}), the  decay rates are $\kappa$ and $\gamma_j$ for the $\hat{a}$ and $\hat{b}_j$ modes respectively and $\mathcal{D}[\hat{A}]\rho = 2 \hat{A} \rho \hat{A}^{\dagger} -\hat{A}^{\dagger} \hat{A} \rho - \rho \hat{A}^{\dagger} \hat{A}$. Furthermore, it is assumed that the losses from the $\hat{b}_j$ modes are negligible on the timescale of interest, $g_j, \kappa \gg\gamma_j$. This assumption requires that the oscillators in the ensemble have high quality factors. In the interaction picture the effective master equation is,
\begin{equation}\label{MEGTc}
\dot{\rho} = -i [ \mathcal{G}_N (\hat{a}^{\dagger} \hat{C}_N + a \hat{C}_N^{\dagger}  ) , \rho] + \frac{\kappa}{2}\mathcal{D}[\hat{a}]\rho, 
\end{equation}
where we have used the approximation, $\gamma_j \approx 0$.

\subsection{Decoherence Free Subspaces}

We now show how decoherence free subspaces with respect to the central $\hat{a}$ mode can be constructed. Using the collective commutation relations, $[\hat{C}_j, \hat{C}^{\dagger}_{j'}  ] = \delta_{j, j'}$, it is easy to show that for any degeneracy vector $\mathbf{d}_L$ the states at the bottom of each ladder do not couple to the $\hat{a}$ mode , i.e.,
\begin{equation}\label{dark}
\hat{C}_N | \mathbf{d}_L, \Phi^0_L \rangle = 0.
\end{equation}
Using this fact it is clear that the state,
\begin{equation}
| \psi_{DFS} \rangle = | 0 \rangle_a \sum_{L, \mathbf{d}_L} c_{(L, \mathbf{d}_L)}  | \mathbf{d}_L, \Phi^0_L \rangle,
\end{equation}
where $\sum_{L, \mathbf{d}_L} | c_{(L, \mathbf{d}_L)}|^2 =1$, under the action of the Hamiltonian (\ref{HGTc}) will not populate the $\hat{a}$ mode. As the central  mode is not populated, the state, $| \psi_{DFS} \rangle$, will not decay into the bath under the evolution of the master equation (\ref{MEGTc}). Therefore the oscillator ensemble prepared in $| \psi_{DFS} \rangle$ is in a decoherence free subspace with respect to the central $\hat{a}$ mode \cite{KnightDFS}.

The collective operators provide a useful method of characterizing all decoherence free subspaces in the system. In the context of superradiance these states are also known as dark states, which are the special case of subradiant states with no radiance. From (\ref{dark}) it is clear that $| \Phi^0_{\beta_k} \rangle$ are dark coherent states.  Dark collective squeezed coherent states can be generated through combinations of $\hat{D}_{k} (\beta)$,  $\hat{S}_{k} (\xi )$ and $ \hat{S}_{k,l} (\xi )$ acting on the multimode vacuum, $  | \mathbf{e}_0, \Phi^0_{0} \rangle $, with $k,l\ne N$. Decoherence free subspaces have wide applications in quantum information \cite{DFSReview, KnightDFS} and in particular the dark collective squeezed coherent states may have use as a quantum memory in continuous variable quantum information \cite{CV_QuantumMemory}.

\subsection{Superradiance Master Equation}

Returning to the master equation (\ref{MEGTc}), we consider preparing the oscillator ensemble in a particular initial state which will then decay into the central $\hat{a}$ mode and rapidly leak out to be observed by a detector. The central mode acts as a common bath for the $N$ surrounding oscillators. In this section we will show how this interaction with the common bath can enhance (superradiance) or reduce (subradiance) emission from the oscillator ensemble.

We now assume the central oscillator is strongly damped, $\kappa\gg \mathcal{G}_N$. This is equivalent to assuming that the $\hat{a}$ mode is always in the ground state \cite{CarmichaelStatisticalMethods2, Mabuchi}. Therefore we can adiabatically eliminate the  $\hat{a}$ modes' degree's of freedom from the master equation (\ref{MEGTc}) \cite{CarmichaelStatisticalMethods2}. After adiabatic elimination, the reduced density matrix for the $N$ modes is given by, 
\begin{equation}\label{rhocN}
\dot{\rho}_b =  \frac{ N \Gamma }{2} \mathcal{D}[\hat{C}_N]\rho_b,  
\end{equation}
where, we have also introduced the effective single oscillator decay rate, $\Gamma = 4 \mathcal{G}^2_N/N \kappa$. This is a useful timescale for the system because for a system of oscillators with identical coupling rates, $g_j = g \, \forall j$, the effective single oscillator decay rate, $\Gamma = 4 N g^2/ N \kappa$, is the same as the decay rate of a single oscillator, $4 g^2/\kappa$. In Appendix \ref{MatrixElementsSRMEAppendix} the density matrix elements of the  master equation (\ref{rhocN}), $P_{(R,L, \mathbf{d}_L)} (t) = \langle \mathbf{d}_L , \Phi^R_L | \rho_b (t) | \mathbf{d}_L , \Phi^R_L \rangle$, are found for an arbitrary pure initial state.

Equation (\ref{rhocN}) is the superradiance master equation  for harmonic oscillators \cite{AgarwalQHOpaper,Agarwal1974}. The superradiance master equation describes the evolution of a system of oscillators which decay into a common Markovian bath. We note that it is not necessary to have a strongly damped mode to create the common bath, as will be discussed in the integrated photonics system in section \ref{imps}. In order to compare systems of oscillators with independent baths to those with a common bath, we state the master equation that describes the decay of an ensemble of oscillators into individual zero temperature heat baths,
\begin{equation}\label{rhocNI}
\dot{\rho}_I = \frac{ \Gamma_I }{2} \sum^N_{j=1}  ( 2 \hat{b}_j \rho_I \hat{b}^{\dagger}_j -\hat{b}^{\dagger}_j \hat{b}_j \rho_I -  \rho_I \hat{b}^{\dagger}_j \hat{b}_j),  
\end{equation}
where, it is assumed each oscillator decays at the rate, $\Gamma_I$. Assuming uniform coupling rates, $g_j =g \, \forall j$, the superradiance master equation can be expanded similarly,
\begin{equation}\label{rhocNU}
\dot{\rho}_b =  \frac{ \Gamma }{2} \sum^N_{i,j=1} ( 2 \hat{b}_j \rho_b \hat{b}^{\dagger}_i -\hat{b}^{\dagger}_i \hat{b}_j \rho_b -  \rho_b \hat{b}^{\dagger}_i \hat{b}_j).  
\end{equation}
The additional, $i\neq j$, terms in the superradiance master equation (\ref{rhocNU}) can enhance or reduce the decay of ensemble states, leading to superradiance or subradiance, respectively.

We now consider the intensity of emission of the oscillators into the common bath. The intensity of emission of quanta is the rate at which the oscillators dissipate quanta,
\begin{equation} \label{IntensityODE}
I_N (t) = - \frac{\partial}{ \partial t} \langle \hat{M} (t) \rangle  .
\end{equation}

Using the superradiance master equation (\ref{rhocN}) it can be shown,
\begin{eqnarray}
\frac{\partial}{ \partial t} \langle \hat{M} (t) \rangle &=& \text{Tr} \{ \hat{M} \dot{\rho}_b (t) \} = -N \Gamma \langle \hat{R} (t) \rangle \\
\frac{\partial}{ \partial t} \langle \hat{R} (t) \rangle &=& \text{Tr} \{ \hat{R} \dot{\rho}_b (t) \} = -N \Gamma \langle \hat{R} (t) \rangle,
\end{eqnarray}
from which we find, $\langle \hat{R} (t) \rangle = \langle \hat{R} (0) \rangle e^{-N \Gamma t}$. After substituting these results into (\ref{IntensityODE}) the intensity can be found, 
\begin{equation} \label{imp2}
I_N (t) = N \Gamma \langle \hat{R} (0) \rangle e^{-N \Gamma t}.
\end{equation}
Hence, the intensity of emission from \textit{any} state is reduced to the problem of calculating $\langle \hat{R} (0)\rangle$. We note the intensity  (\ref{imp2}) is consistent with the phase space approach of Agarwal \cite{AgarwalQHOpaper} after a suitable parameter substitution.

It can also be shown from (\ref{rhocN}) that $\langle \hat{L} \rangle$ is conserved,
\begin{equation}
\frac{\partial}{ \partial t} \langle \hat{L} (t) \rangle = \text{Tr} \{ \hat{L} \dot{\rho}_b  \} = 0.
\end{equation}
Therefore we can associate the following physical meanings with the operators: $\langle \hat{M} \rangle$ is the number of quanta in the system,  $\langle \hat{L} \rangle$ is the number of trapped (dark) quanta in the system and  $\langle \hat{R} \rangle$ is the number of bright quanta that is free to radiate. It is interesting to note that the intensity is proportional to the initial number of bright quanta in the system, $\langle \hat{R} (0) \rangle$, not the initial energy in the system, $\langle \hat{M} (0) \rangle=\langle \hat{L} (0) \rangle+\langle \hat{R} (0) \rangle$.

\section{Criterion for Superradiance in Harmonic Oscillators \label{SRcriterionSection}}

In this section we classify the initial states in the ensemble that lead to, respectively, superradiant, normal or subradiant emission. In atomic ensembles the definition is clear cut: normal radiance has an intensity proportional to the number of atoms, $N$, superradiant states have an intensity greater than $N$ and subradiant states have emission intensity less than $N$. Furthermore, the maximum intensity of emission from the superradiant states scales with $N^2$ and the minimum intensity from subradiant dark states  is zero. For ensembles of harmonic oscillators we develop a similar criterion.

To resolve the classification of initial states for harmonic oscillators  we expand the intensity (\ref{imp2}),
\begin{eqnarray}
I_N (t) &=& N \Gamma e^{-N \Gamma t}  \langle \hat{R} (0) \rangle \nonumber \\
&=&  \frac{N \Gamma}{\mathcal{G}^2_N} e^{-N \Gamma t} \sum^N_{i,j=1} g_i g_j \langle b^{\dagger}_i (0) \hat{b}_j (0) \rangle.
\end{eqnarray}
It is clear that the intensity depends on the initial correlations between the oscillators. Normal incoherent, uncorrelated radiance from an ensemble of independent oscillators with independent baths have, $\langle \hat{b}^{\dagger}_i (0) \hat{b}_j (0) \rangle = 0$ for all $i\neq j$ \cite{Agarwal1974}. On the other hand, coherent correlated radiation results from the correlations within an ensemble of harmonic oscillators as they interact with a common bath, i.e. $\langle \hat{b}^{\dagger}_i (0) \hat{b}_j (0) \rangle \neq 0$ for some $i\neq j$. Therefore, assuming $N>1$, we can separate the intensity (\ref{imp2}) into two parts \cite{Agarwal1974},
\begin{eqnarray}
I_N (t) =  I^{U}_N (t) + I^{C}_N (t), 
\end{eqnarray}
where the uncorrelated and correlated parts are, respectively,
\begin{eqnarray}
 I^{U}_N (t) &=&  \frac{N \Gamma}{\mathcal{G}^2_N}  e^{-N \Gamma t} \sum^N_{j=1}  g^2_j\langle  \hat{b}^{\dagger}_j (0) \hat{b}_j (0) \rangle, \label{IncohIntensity} \\
 I^{C}_N (t) &=& \frac{N \Gamma}{\mathcal{G}^2_N} e^{-N \Gamma t} \sum^N_{i\neq j} g_i g_j \langle \hat{b}^{\dagger}_i (0) \hat{b}_j (0) \rangle. \label{cohIntensity}
\end{eqnarray}
We see that normal radiance comes from the uncorrelated part and superradiance or subradiance from the correlated part. The correlated part is only nonzero for states where there exists $\langle \hat{b}^{\dagger}_i (0) \hat{b}_j (0) \rangle \neq 0$ for at least a single pair of modes, $i \neq j$. Therefore, we can classify initial states with a \textit{superradiance criterion}: superradiance occurs for  $I^{C}_N (t)>0$, normal radiance occurs when  $I^{C}_N (t)=0$, and subradiance occurs when $I^{C}_N (t)<0$. 

Another useful quantity for classifying the radiance of initial states is the fraction of the total energy that is trapped in the system after emission. The fraction of energy that is dark (trapped) is given by $F = \langle \hat{L} (0)\rangle/\langle \hat{M}(0)\rangle = 1- \langle \hat{R} (0)\rangle/\langle \hat{M} (0)\rangle$, where $F\in [0,1]$. When $F=0$ all quanta are emitted from the ensemble, whereas when $F=1$ no quanta are emitted. Using the superradiance criterion, we find that a normal state has the fraction,
\begin{subequations}
\label{SRCriterion}
\begin{equation} \label{FNSRcond}
F_N = 1- \frac{ \sum^N_{j=1} g^2_j \langle  \hat{b}^{\dagger}_j (0) \hat{b}_j (0) \rangle }{ \mathcal{G}^2_N  \sum^N_{j=1}  \langle  \hat{b}^{\dagger}_j (0) \hat{b}_j (0) \rangle } \Leftrightarrow \text{Normal radiance},
\end{equation} 
and for any initial state the conditions hold,
\begin{eqnarray}
F &<& F_N \Leftrightarrow \text{Superradiance}, \label{SRCriterionSR} \\
F &>& F_N \Leftrightarrow \text{Subradiance}.\label{SRCriterionSubR}
\end{eqnarray}
\end{subequations}
In the special case of uniform coupling rates, $g_j =g \, \forall j$, these conditions simplify considerably,
\begin{subequations}
\label{UniformgSRcond}
\begin{eqnarray}
F_N &=& 1-\frac{1}{N} \Leftrightarrow \text{Normal radiance},  \\
F &<& 1-\frac{1}{N} \Leftrightarrow \text{Superradiance},  \\
F &>& 1-\frac{1}{N} \Leftrightarrow \text{Subradiance}.
\end{eqnarray}
\end{subequations}
These conditions provide a very simple method of classifying all initial states. It is interesting to note that for normal states with large $N$, most of the energy in the system is trapped.

The fraction of dark energy monotonically decreases with the intensity, i.e, if two states, $a$ and $b$, have the same energy, $\langle \hat{M} (0)\rangle_a = \langle \hat{M} (0)\rangle_b$ then,
\begin{equation}
I^a_{N} > I^b_{N} \Leftrightarrow  F_a < F_b
\end{equation}
where $I^i_{N}$ is the initial intensity and $F_i$ is the fraction of energy trapped of state $i=a,b$. Therefore, the maximum radiance from an ensemble of $N$ oscillators with $\langle \hat{M}\rangle$ quanta occurs when $F=0$. Similarly the minimum intensity occurs for states with  $F=1 \Leftrightarrow I_{N} (t) =0  \Leftrightarrow \langle \hat{R} (0) \rangle=0  $. The fraction of dark energy is therefore a useful measure of the radiance from the ensemble.

\section{Superradiance from Various Initial States \label{SRIc}}

In this section we compare the emission dynamics for a range of interesting initial states. These states include bosonic Dicke basis states, multimode Fock states, thermal states, product single mode squeezed coherent states and collective mode squeezed coherent states.  We determine parameter regimes where these states are superradiant according to the superradiance criterion. Several properties of interest are also calculated including the emission intensity, fraction of dark quanta and two-time correlation functions. 

Two-mode correlations are an important observable for superradiance as these correlations are the cause of the enhanced emission intensity in superradiance. These correlations could be used to differentiate between states that have the same emission intensity. Two-time correlation functions between two modes can be calculated using the the quantum regression theorem,
\begin{eqnarray}
c_{i,j} (t , 0) &=& \langle \hat{b}^{\dagger}_i (t) \hat{b}_j (0) \rangle - \langle \hat{b}^{\dagger}_i (t) \rangle \langle \hat{b}_j (0) \rangle \\
&=& -\frac{g_i}{ \mathcal{G}_N} (1- e^{-N \Gamma t}) \Big( \langle \hat{C}^{\dagger}_N (0) \hat{b}_j (0)  \rangle  \nonumber \\
&-&  \langle \hat{C}^{\dagger}_N (0) \rangle   \langle \hat{b}_j (0)  \rangle  \Big) \nonumber \\
&+& \langle \hat{b}^{\dagger}_i (0) \hat{b}_j (0) \rangle - \langle \hat{b}^{\dagger}_i (0) \rangle \langle \hat{b}_j (0) \rangle.
\end{eqnarray}
We demonstrate the applicability of the bosonic Dicke states for calculating these quantities on several interesting initial states.

\subsection{Bosonic Dicke State \label{Singlecollective}}

We start by considering the simplest initial state in the bosonic Dicke basis,
$|\mathbf{d}_L, \Phi^R_L\rangle$. As discussed previously,  this state is difficult to prepare in general, however, for a small number of oscillators and few total quanta ($\langle \hat{M} (0)\rangle<3$) the lower lying states of the bosonic Dicke basis are currently within reach experimentally. From the superradiance master equation (\ref{rhocN}) it can be seen that $|\mathbf{d}_L, \Phi^R_L\rangle$ will decay down the $L$-th ladder and as $t \rightarrow \infty$ all population will reside in $|\mathbf{d}_L, \Phi^0_L\rangle$. The intensity from the  $|\mathbf{d}_L, \Phi^R_L\rangle$ state is,
\begin{equation}\label{IcollBasisstate}
I_N (t) = R N \Gamma  e^{-N \Gamma t},
\end{equation}
an exponential decay proportional to the number of bright quanta in the system.

\begin{figure}[t!]
\centering
\subfigure[]{
{\includegraphics[scale=0.0709]{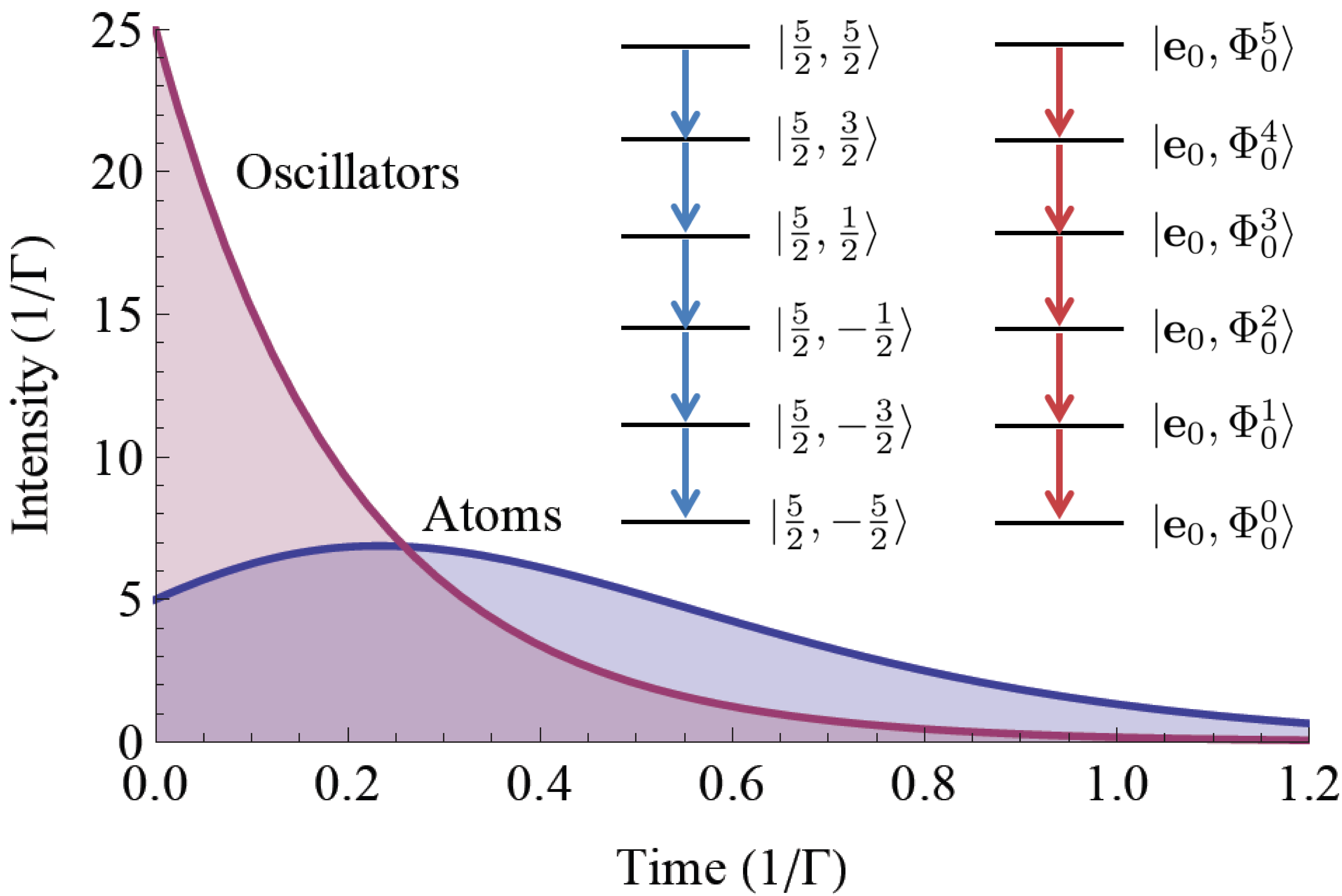}\label{FiveLevelDecay}}
}
\subfigure[]{
{\includegraphics[scale=0.7]{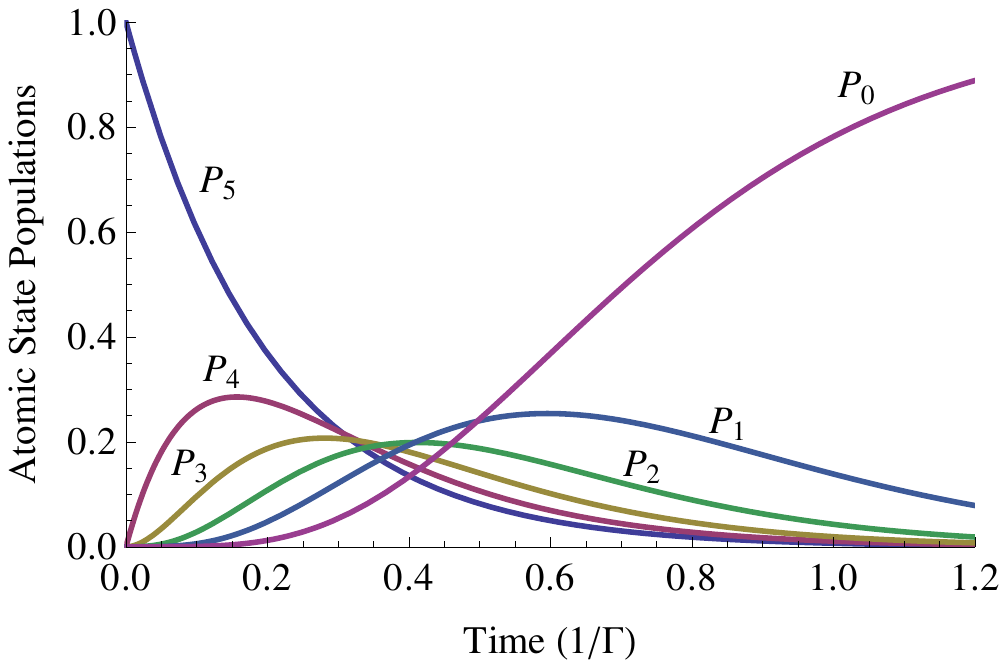} \label{POPDiagramA}}
}
\subfigure[]{
{\includegraphics[scale=0.7]{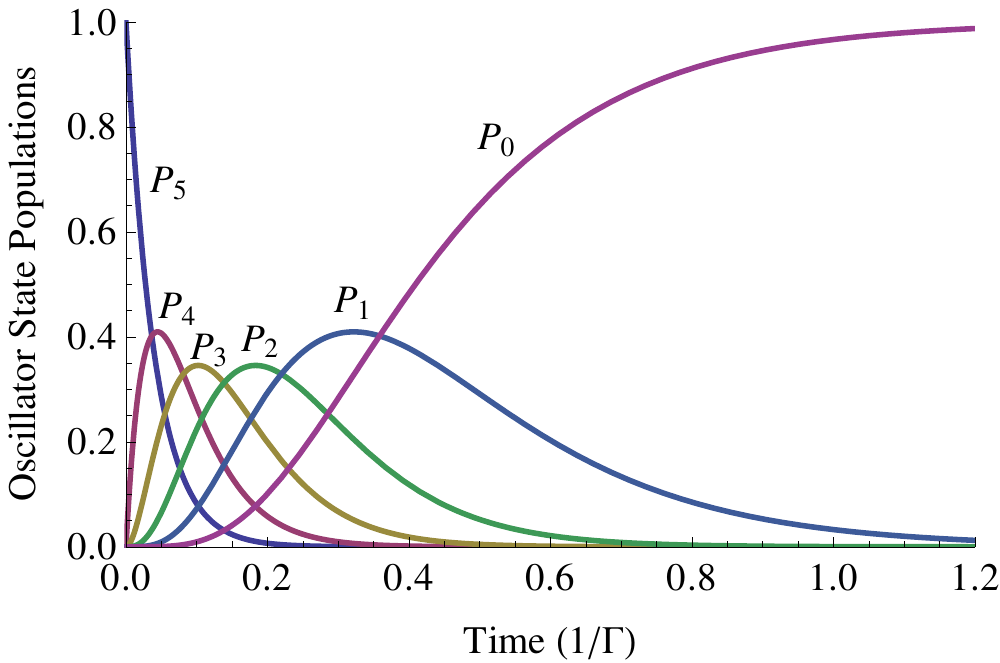}  \label{POPDiagramF}}
}
\caption[]{Comparison of the emission from $N=5$ atoms prepared in the  state $|\frac{5}{2}, \frac{5}{2} \rangle$ and $N=5$ oscillators prepared in $| \mathbf{e}_0 ,\Phi^5_0\rangle$. \textit{Intensity of emission}: (a)  the oscillators (\ref{IcollBasisstate}) have a significantly larger emission intensity than the atoms (\ref{IntensitySingleAtomicDickeState5}) despite each state containing five quanta, $\langle \frac{5}{2}, \frac{5}{2} | J_z |\frac{5}{2}, \frac{5}{2} \rangle = \langle \Phi^5_0| \hat{M}| \Phi^5_0 \rangle = 5$. \textit{State populations}: (b)   The populations of the atomic Dicke states during emission, $P_K = \langle \frac{5}{2},K- \frac{5}{2} | \rho_q |\frac{5}{2}, K-\frac{5}{2} \rangle $ from (\ref{DickeStatePops}). (c) The populations of the bosonic Dicke states during emission $P_K = \langle \Phi^K_0| \rho_b | \Phi^K_0 \rangle $ from (\ref{PopSingleBosonicDicke}). It is assumed all atoms and  oscillators couple to the strongly damped mode at the same rate $g$.}
\end{figure}

Now that we have an expression for the intensity from any bosonic Dicke basis state we compare this to the intensity from an ensemble of atoms. We consider the emission from five atoms initially in the state $|\frac{5}{2}, \frac{5}{2}\rangle= |e,e,e,e,e \rangle$ and compare this to the emission from five oscillators in the equivalent state, $| \mathbf{e}_0, \Phi^5_0\rangle$. In the atomic (bosonic) Dicke basis, this state is the fifth state above the ground state, $|\frac{5}{2}, -\frac{5}{2}\rangle=|g,g,g,g,g\rangle$ ($| \mathbf{e}_0, \Phi^0_0\rangle = |0,0,0,0,0\rangle$) under the action of the collective raising operator $J_+$ ($\hat{C}^{\dagger}_N$). Fig.~\ref{FiveLevelDecay} shows the intensities for these initial states. As expected, the oscillators have a significantly larger intensity and decay much more rapidly than the atoms.
This is because the oscillator states $|\mathbf{e}_0, \Phi^K_0\rangle$ with $K=2$-$5$, have a significantly larger decay rate $\propto 5 K$, than the atomic state $|\frac{5}{2}, K-\frac{5}{2}\rangle$, which decay at the rate $\propto K(6-K)$. The populations of the atomic and oscillator states are also shown in  Fig.~\ref{POPDiagramA} and Fig.~\ref{POPDiagramF}. It can be seen that the atomic states have a much larger spread in populations than the oscillators.

Returning to the analysis of the emission of the bosonic Dicke basis state, $|\mathbf{d}_L, \Phi^R_L\rangle$, we find that the fraction of dark energy is $F = L/(L+R)$. From this fraction we see that the higher the state is on the ladder, $R$, the smaller the amount of trapped photons. On the other hand the bottom of each ladder has $R=0$ and $F=1$ and therefore all energy is trapped. Assuming uniform coupling rates, using the superradiance criterion (\ref{UniformgSRcond}) we find the state is superradiant for $L< R(N-1)$. 

The two-time correlation functions, $c_{i,j} (t , 0)$, are difficult to compute in general as they are strongly dependent on the two modes ($i,j$) and the particular initial state $|\mathbf{d}_L, \Phi^R_L\rangle$ considered. However, for states on the first ladder, $|\mathbf{e}_0, \Phi^R_0\rangle$, the correlation function has a particularly simple form,
\begin{equation}\label{collBasisStatecorr}
c_{i,j} (t , 0)  = R \frac{ g_i g_j }{\mathcal{G}^2_N} e^{-N \Gamma t},
\end{equation}
indicating that the correlation between the oscillators exponentially decays over time.

\subsection{Multimode Fock State}
We now consider preparing the oscillators in the multimode Fock state, 
\begin{equation} \label{MultimodeFockState}
|n_1, n_2, \dots, n_N\rangle = \left( \prod^N_{j=1} \frac{(\hat{b}^{\dagger}_j)^{n_j}}{\sqrt{n_j!}}  \right) | 0 \rangle^{\bigotimes N}.
\end{equation}
As the oscillators are initially uncorrelated, we expect very different behavior from the multimode Fock state than from the bosonic Dicke basis state. The multimode Fock state is inconvenient to express in the bosonic Dicke basis, however it is still possible to calculate the intensity using (\ref{imp2}),
\begin{equation} \label{ImultiFock}
I_N (t) =  N \Gamma \frac{e^{-N \Gamma t}}{\mathcal{G}^2_N} \sum^N_{j=1} g^2_j n_j.
\end{equation}

The fraction of dark quanta is,
\begin{equation}
F = 1- \frac{ \sum^N_{j=1} g^2_j n_j }{\mathcal{G}^2_N \sum^N_{j=1} n_j } = F_N,
\end{equation}
indicating that the state has normal incoherent radiance. It is surprising that for uniform coupling rates ($g_j=g \, \forall j$) the fraction is, $F=1-\frac{1}{N}$, which is independent of the number of photons in each oscillator, $n_j$. Furthermore, for a large number of oscillators most of the excitation is dark as $F=1-\frac{1}{N} \rightarrow 1$. 

It is interesting to compare the intensity of the multimode Fock state (\ref{ImultiFock}) to that of the bosonic Dicke state (\ref{IcollBasisstate}). Assuming we have $K$ quanta in the initial state and uniform coupling rates ($g_j=g \, \forall j$), the multimode Fock state has the intensity $I_N (t) = K \Gamma e^{-N \Gamma t}$, whereas the bosonic Dicke basis state, $| \mathbf{e}_0, \Phi^K_0 \rangle $, has $I_N (t) = K N \Gamma e^{-N \Gamma t}$, an $N$-fold enhancement one would expect from superradiance.

The two-time correlations between the modes are,
\begin{equation} \label{corrFock}
c_{i,j} (t , 0)  = - \frac{ g_i g_j }{\mathcal{G}^2_N} (1- e^{-N \Gamma t}) n_j + \delta_{i,j} n_j,
\end{equation}
where, $\delta_{i,j}$ is the Kronecker delta. For $i\neq j$ the correlation is initially zero and becomes more negative over time due to the decay inducing correlations between the modes. The initial unentangled product state becomes increasingly entangled as it decays. For equal coupling rates the intensity (\ref{ImultiFock}) and two mode correlations (\ref{corrFock}) are consistent with \cite{AgarwalQHOpaper}.

\subsection{Incoherent Mixtures and Thermal states}
In this section we consider a product of single oscillator incoherent mixtures. These mixtures include a product of thermal states in each oscillator, which are particularly simple to prepare. Due to their classical nature, we do not expect that these states will exhibit any superradiant phenomena. 

A product of single oscillator incoherent mixtures  is described by the density matrix,
\begin{eqnarray}
\rho(0) = \bigotimes^{N}_{j=1} \rho_j(0), \text{with,} \label{IncohMixDensityMat}\\
\rho_j(0) = \sum^{\infty}_{n=0} P^j_n | n \rangle_j \langle n |,
\end{eqnarray}
where, $\sum^{\infty}_{n=0} P^j_n =1$ and we denote the average photon number in each oscillator by $\bar{n}_j = \sum^{\infty}_{n=0} n P^j_n $. For example a thermal state in the $j$-th oscillator has,
\begin{equation}
P^j_n = \frac{ (\bar{n}_j)^{n} }{ (\bar{n}_j+1)^{n+1}  },
\end{equation}
where, $ \bar{n}_j= 1/(e^{\hbar \omega /k T_j} -1)$ is the average thermal photon number of the  $j$-th oscillator at temperature, $T_j$. Furthermore the multimode Fock state considered in (\ref{MultimodeFockState}) is also an incoherent mixture with $P^j_n = \delta_{n,n_j }$.

As the density matrix (\ref{IncohMixDensityMat}) factorizes it is clear that  $\langle  b^{\dagger}_i (0) \hat{b}_j (0) \rangle =0 \, \forall i\neq j$. This implies that the correlated part of the intensity (\ref{cohIntensity}) is zero, indicating that an incoherent mixture has normal radiance. The fraction of dark quanta,
\begin{equation} \label{Fincoh}
F = 1- \frac{ \sum^N_{j=1} g^2_j \bar{n}_j }{\mathcal{G}^2_N \sum^N_{j=1} \bar{n}_j } = F_N,
\end{equation}
supports this conclusion. Therefore, the emission from a system of oscillators in a thermal state is normal. Although the oscillators are initially uncorrelated, the emission into the common bath induces correlations as can be seen from the two-time correlation function,
\begin{equation} \label{corrInchohMix}
c_{i,j} (t , 0)  = - \frac{ g_i g_j }{\mathcal{G}^2_N} (1- e^{-N \Gamma t}) \bar{n}_j + \delta_{i,j} \bar{n}_j.
\end{equation}

Finally, the intensity of emission from an incoherent mixture is similar to the multimode Fock state,
\begin{equation} \label{Iincoh}
I_N (t) =  N \Gamma \frac{e^{-N \Gamma t}}{\mathcal{G}^2_N} \sum^N_{j=1} g^2_j \bar{n}_j.
\end{equation}
For uniform coupling rates, $g_j =g \, \forall j$, and average photon numbers, $\bar{n}_j = \bar{n} \, \forall j$, the intensity simplifies to,
\begin{equation}
I^U_N (t) =  \bar{n} N \Gamma e^{-N \Gamma t}.
\end{equation}
We see that the incoherent mixture has the intensity one would expect from normal radiance, $I^U_N (t) \propto  N$.

\subsection{ Squeezed Coherent States \label{ProSqcoh}}
%Product Squeezed Coherent States in the Fock Basis
Another state which is straightforward to prepare is a product of single mode coherent states. This state is of particular interest because of the close connection between coherent states and classical harmonic oscillators. The emission from an ensemble of classical oscillators with random phase scales with $I_N \propto N$, whereas for classical oscillators that are in-phase the emission scales as, $I_N \propto N^2$. The in-phase intensity is clearly reminiscent of superradiance. 

In addition to coherent states, we also consider the possibility of single mode squeezing \footnote[5]{Two mode squeezing in the Fock basis has similar results.} in each oscillator, as each mode can be prepared independently. Therefore, we consider the initial state where the oscillators are prepared in a product of single mode squeezed coherent states,
\begin{eqnarray} \label{SQCOHSTATEDEF}
\! \! \! \! \! \! \! \! \! \! \! \! \! \! \! \! \! 
|\alpha_1, \xi_1 \rangle_1 |\alpha_2, \xi_2 \rangle_2 \dots | \alpha_N, \xi_N \rangle_N \nonumber\\
\! \! \! \! \! \! \! \! \! \! \! \! \! \! \! \! \!
= \prod^N_{j=1} e^{\alpha_j \hat{b}_j^{\dagger} - \alpha_j^* \hat{b}_j} e^{\frac{1}{2} ( \xi_j^* \hat{b}^2_j - \xi_j (\hat{b}_j^{\dagger})^2 )} |0 \rangle^{\bigotimes N},
\end{eqnarray}
where, $\xi_j = r_j e^{i \theta_j}$. The intensity of emission for this state is,
\begin{equation}\label{IntensityProdSqucoh}
I_N (t) =  \frac{N \Gamma e^{-N \Gamma t}}{ \mathcal{G}^2_N } \left(  |\sum^N_{j=1} g_j \alpha_j  |^2 + \sum^N_{j=1}  g^2_j \sinh^2 r_j      \right).
\end{equation}
The most important feature of the intensity is that the amplitudes, $\alpha_j$, add coherently. This feature is most apparent for the special case of uniform coupling rates, displacements and squeeze amplitudes ($g_j =g, \alpha_j=\alpha,r_j=r   \, \forall j$),
\begin{equation} \label{IntensSqcohU}
I^U_N (t) = \Gamma (N^2 |\alpha|^2 + N \sinh^2 r )e^{-N \Gamma t}.
\end{equation}
We see that the intensity has the $I^U_N (t) \propto N^2$ characteristic of atomic superradiance. However, the intensity from the squeezing contribution only scales linearly with the number of oscillators, a characteristic of normal radiance. In fact for squeezed vacuum states ($\alpha = 0$) the intensity is $I^U_N (t) \propto N$, whereas for coherent states ($r=0$), $I^U_N (t) \propto N^2$. 

In order to classify the radiance of the state (\ref{SQCOHSTATEDEF}), we find the fraction of dark quanta, 
\begin{equation}\label{Fprodsqcoh}
F = 1- \frac{  |\sum^N_{j=1} g_j \alpha_j  |^2 + \sum^N_{j=1}  g^2_j \sinh^2 r_j   }{ \mathcal{G}^2_N  \sum^N_{j=1} \left( |\alpha_j  |^2 +  \sinh^2 r_j     \right)}.
\end{equation}
Using this fraction and the superradiance criterion we find that the state is superradiant if \footnote[6]{A product of single mode displaced thermal states is also superradiant under the condition (\ref{ProdcohStateSRcond}). },
\begin{equation}\label{ProdcohStateSRcond}
 |\sum^N_{j=1} g_j \alpha_j  |^2 > \sum^N_{j=1}  g^2_j | \alpha_j  |^2.
\end{equation}
Therefore, the superradiance of the state is independent of the squeezing amplitudes, $r_j$. If each oscillator amplitude has a random phase, $|\sum^N_{j=1} g_j \alpha_j  |^2 = \sum^N_{j=1}  g^2_j | \alpha_j  |^2$, and it can be shown that the state has normal radiance. This phenomenon is analogous to classical oscillators.

\begin{figure}[t!]
\centering
{\includegraphics[scale=0.7]{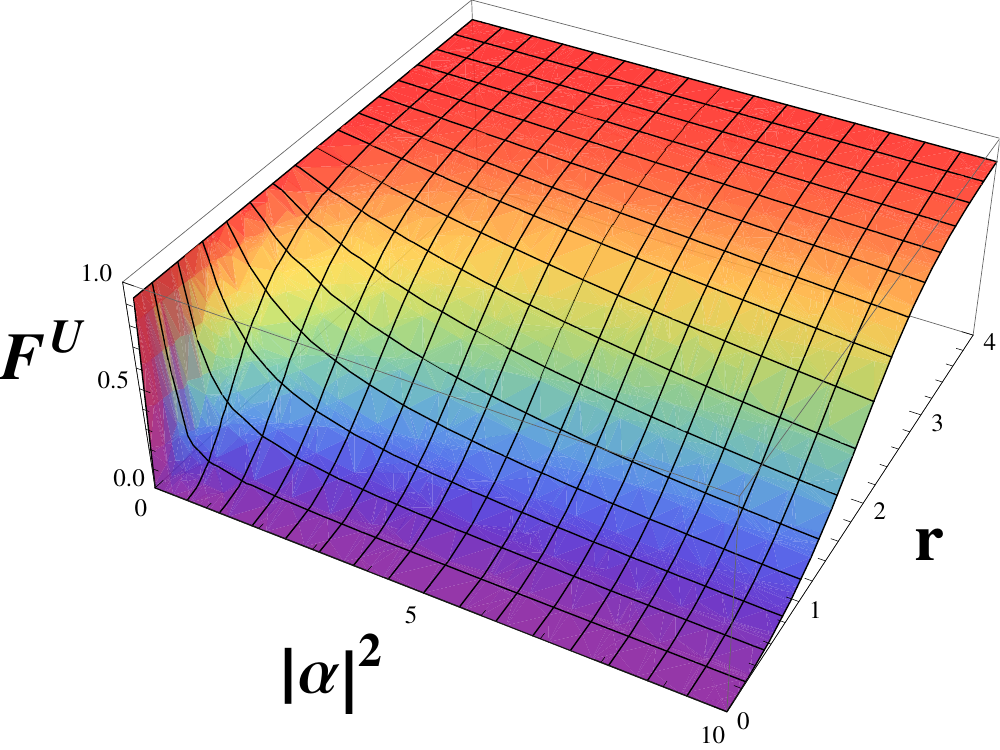} }
\caption[]{Fraction of trapped quanta from $N=10$ oscillators prepared in a product squeezed coherent state $|\alpha, r e^{i \theta} \rangle^{\bigotimes 10}$ (\ref{FUProdSqcoh}). It is assumed all  coupling rates, displacements and squeeze amplitudes are identical, $g_j =g, \alpha_j=\alpha,r_j=r \, \forall j$. Normal radiance occurs for $F^U = F_N = 1-\frac{1}{10} =0.9$, superradiance occurs for $F^U < 0.9$. We see that large squeezing amplitudes trap dark quanta in the system and the maximum radiance occurs for coherent states ($r=0$). }
\label{FRacDiagram} 
\end{figure}

In the special case of uniform coupling rates, displacements and squeeze amplitudes ($g_j =g, \alpha_j=\alpha,r_j=r \, \forall j$), the fraction of trapped energy becomes,
\begin{equation} \label{FUProdSqcoh}
F^U = 1- \frac{|\alpha|^2 + \frac{1}{N}  \sinh^2 r }{|\alpha|^2 +  \sinh^2 r },
\end{equation}
which is shown in Fig.~\ref{FRacDiagram}. It is clear that $F^U \in [0, 1- \frac{1}{N}]$, with $F^U=0$ for $r=0$ and $F^U=1-\frac{1}{N} = F_N$ when $\alpha=0$. Therefore for $|\alpha|>0$ the state is always superradiant, reaching the maximum intensity of emission when there is no squeezing, $r=0$. Furthermore, single mode squeezing has the effect of trapping photons in the system as $r>0 \Rightarrow F^U >0$. Finally, the squeezed vacuum state, $\alpha=0$, exhibits normal radiance.

The two-time correlation function between the modes is,
\begin{equation} \label{corrsqucohprod}
c_{i,j} (t , 0)  = - \frac{ g_i g_j }{\mathcal{G}^2_N} (1- e^{-N \Gamma t}) \sinh^2 r_j + \delta_{i,j} \sinh^2 r_j.
\end{equation}
This coincides with the incoherent mixture correlation function (\ref{corrInchohMix}) after the replacement, $\sinh^2 r_j \rightarrow \bar{n}_j$. We note that for no squeezing, $r_j =0$, the correlation function is zero, $c_{i,j} (t , 0)  = 0$. Therefore correlations between the modes are not induced during the decay of product coherent states. This is surprising because in atomic superradiance, superradiant states are highly entangled and there is a large correlation between emitters during emission.

Summarizing the above results, we see that a product of single mode coherent states can be superradiant (\ref{ProdcohStateSRcond}), whereas a product of single mode squeezed states possesses many of the properties of an incoherent mixture. For example, consider the state with no displacement ($\alpha_j=0$), $|\xi_1 \rangle_1 |\xi_2 \rangle_2 \dots | \xi_N \rangle_N$. The average photon number in each mode, $\bar{n}_j$, is $ \langle \hat{b}^{\dagger}_j \hat{b}_j \rangle = \sinh^2 r_j$. Under the replacement $\sinh^2 r_j \rightarrow \bar{n}_j$, we find that the intensity (\ref{IntensityProdSqucoh}), fraction of trapped quanta (\ref{Fprodsqcoh}) and two-time correlation function (\ref{corrsqucohprod}) all reduce to the corresponding equations for the incoherent mixture, (\ref{Iincoh}), (\ref{Fincoh}) and (\ref{corrInchohMix}) respectively. Furthermore from the fraction of trapped quanta we see that the radiance is normal. We conclude that single mode squeezing does not lead to superradiant behavior.

\subsection{Collective Squeezed Coherent States}

\begin{table*}
\caption{\label{table1} Collective Squeezed Coherent States}
\begin{ruledtabular}
\begin{tabular}{cccc}
State & Intensity ($I_N (t)$) & Condition for Superradiance & Fraction of Dark Quanta ($F$)\\
$ | \Phi^{\alpha}_L \rangle =\hat{D}_{N} (\alpha)  | \mathbf{d}_L, \Phi^{0}_L \rangle$ & $ N \Gamma |\alpha|^2 e^{-N \Gamma t}$ &  $|\alpha|^2(N-1)>L$ \footnote[1]{Assuming uniform coupling rates, $g_j=g \; \forall j$ .} & $ \frac{L}{(L+|\alpha|^2)}$\\
$ |  \Phi^R_{\beta_k} \rangle  = \hat{D}_{k} (\beta)  | \mathbf{e}_0, \Phi^R_{0} \rangle$\footnote[2]{$(k\neq N)$} & $R N \Gamma  e^{-N \Gamma t}$ & $ R(N-1)> |\beta|^2$ $^{a}$ & $\frac{|\beta|^2}{(|\beta|^2+R)}$ 
\\
$\prod^N_{j=1} \hat{D}_{j} (\alpha_j) \hat{S}_{j} (\xi_j) |\mathbf{e}_0,\Phi^0_0 \rangle$ & $ N \Gamma ( |\alpha_N|^2 + \sinh^2 r_N ) e^{-N \Gamma t}$ & $ |\alpha_N|^2 + \sinh^2 r_N  >\frac{1}{\mathcal{G}^2_N}\sum^{N}_{j=1} g^2_j \bar{n}_j$ \footnote[3]{$\bar{n}_j = |\alpha_j|^2 + \sinh^2 r_j $} & $1- \frac{ |\alpha_N|^2 + \sinh^2 r_N  }{ \sum^{N}_{j=1} \bar{n}_j  }$\\
\end{tabular}
\end{ruledtabular}
\end{table*}

In this section we consider preparing the oscillator ensemble in the collective squeezed coherent states from section \ref{cohSec}. The intensity, fraction of dark quanta and conditions for superradiance are listed in Table \ref{table1} for three particular initial states: a coherent state in the $N$-th collective mode, $ | \Phi^{\alpha}_L \rangle$, a coherent state in the $k$-th collective mode, $ |  \Phi^R_{\beta_k} \rangle$,  ($k\neq N$) and a product of collective squeezed coherent states.  With reference to Fig.~\ref{EnergyLevels},   
$ | \Phi^{\alpha}_L \rangle$ are coherent states in the bosonic Dicke basis that occupy states vertically on a single ladder denoted by, $\mathbf{d}_L$. On the other hand the coherent states, $|  \Phi^R_{\beta_k} \rangle$, populate bosonic Dicke basis states along a diagonal: starting from the `ground' state $| \mathbf{e}_0, \Phi^R_{0} \rangle$, the state occupies one position on the $R$-th rung on each ladder labeled by $\mathbf{d}_L = L \mathbf{e}_k$. Furthermore the product collective squeezed coherent states in Table \ref{table1} with $\alpha_j \neq 0 \, \forall j$ will have some population on all rungs of all ladders. 

We first discuss the collective coherent state, $ | \Phi^{\alpha}_L \rangle$. As the decay does not affect the ladder label, evolution under the superradiance master equation (\ref{rhocN}) will drive the population down the ladder to the dark state $| \mathbf{d}_L, \Phi^{0}_L \rangle$. Therefore the intensity is  independent of the undisplaced initial state $| \mathbf{d}_L, \Phi^{0}_L \rangle$ \footnote[7]{We note that for equal coupling rates ($g_j = g\, \forall j$), $| \alpha \rangle^{\bigotimes N} = | \Phi^{ \sqrt{N} \alpha}_L \rangle =\hat{D}_{N} (\sqrt{N} \alpha)  | \mathbf{d}_L, \Phi^{0}_L \rangle$ and the intensity (\ref{IntensSqcohU}) with $r=0$ coincides with the collective coherent state $| \Phi^{ \sqrt{N} \alpha}_L \rangle$ in Table \ref{table1}. }. Superradiance only occurs if $|\alpha|^2$ is large with respect to the number of dark quanta, $L$. Two-time correlation functions between the modes are in general difficult to calculate as it depends on the particular degeneracy vector, $\mathbf{d}_L$. However in the simplest case, 
\begin{equation}
 | \Phi^{\alpha}_0 \rangle = \bigotimes^N_{j=1} | \alpha \frac{g_j}{\mathcal{G}_N} \rangle_j, 
\end{equation}
the state is a special case of the product coherent state in section \ref{ProSqcoh}, with $\alpha_j =  \alpha \frac{g_j}{\mathcal{G}_N}$ and $\xi_j =0$. From (\ref{corrsqucohprod}) we find that no correlations are induced during emission as $c_{i,j} (t , 0)  = 0$.

Next we consider the collective coherent state $ |  \Phi^R_{\beta_k} \rangle$. As the displacement operator $\hat{D}_k (\beta)$ does not add any bright quanta to the system the intensity coincides with that of the bosonic Dicke basis state (\ref{IcollBasisstate}). This state is only superradiant for small displacements, $\beta$. Similar to the state $ | \Phi^{\alpha}_L \rangle$ we note that the two-time correlation functions depend on the specific degeneracy vector, $\mathbf{d}_L$, however, for the simplest case, $ |  \Phi^0_{\beta_k} \rangle $, it can be shown that $c_{i,j} (t , 0)  = 0$.

Finally, we consider the product collective coherent state in Table \ref{table1}. As only the $N$-th collective mode decays, displacement and squeezing of the other modes has no effect on the intensity. It is interesting to note that for uniform displacements and squeezing, $\alpha_j = \alpha, r_j =r \, \forall j$, the state has normal radiance, $F=1-\frac{1}{N}=F_N$.  As discussed above the two-time correlation function is diffucult to compute, however for the $N$-th collective mode squeezed vacuum (\ref{SqucollVac}), $| \Phi^{\xi}_0 \rangle$, it can be found,
\begin{equation}
c_{i,j} (t,0) = \frac{g_i g_j}{\mathcal{G}^2_N} e^{-N \Gamma t} \sinh^2 r_N.
\end{equation}

The  $N$-th collective mode squeezed vacuum, $| \Phi^{\xi}_0 \rangle$, possesses many of the properties of the superradiant bosonic basis state, $ | \mathbf{e}_0, \Phi^R_0 \rangle$; from Table \ref{table1} the intensity, fraction of trapped quanta and  two-time correlation function of $| \Phi^{\xi}_0 \rangle$, are identical to those of $ | \mathbf{e}_0, \Phi^R_0 \rangle$, (\ref{IcollBasisstate}), $F=0$, (\ref{collBasisStatecorr}) respectively, under the replacement $\sinh^2 r_N \rightarrow R$.  This is in contrast to the single mode squeezed states in section \ref{ProSqcoh}, which were similar to normal radiant incoherent mixtures. We conclude that squeezing of the $N$-th collective mode can lead to superradiant behavior.

\section{Implementations \label{imps}}

In this section we examine two experimental implementations which could be used to observe superradiance of harmonic oscillators from a variety of initial states. We consider both optical and microwave implementations. In the optical regime, we discuss direct write integrated photonic waveguide arrays  in integrated photonics  \cite{DirectWriteThorne, OBrienIPReview, ObrienNatureReview, LonghiQOAnalogiesReview}. In the microwave regime  setups based on the circuit QED architecture are considered including stripline \cite{Blais2004} and lumped element resonators \cite{LumpedElement}  and the transmon superconducting qubit tuned to the harmonic regime \cite{TransmonPaper}. In each system initial state preparation is considered and it is shown how the superradiance master equation (\ref{rhocN})  describes the dynamics under a specific parameter regime. 

In choosing an experimental implementation there are three main requirements: high fidelity state preparation of a variety of states, small losses and high fidelity measurement of the superradiant emission. Two mode correlation measurements are also desirable to measure the interaction between the modes in the ensemble during superradiance. Firstly, low fidelity state preparation can lead to transitions along undesired emission pathways, which can reduce the intensity of emission by populating dark states. Secondly, losses such as those from low-$Q$ resonators are a form of non-radiative decay and also lead to undesired transition pathways, which in turn results in reduced intensity. Lastly, low efficiency detectors and dark counts can also reduce and/or distort the measured emission intensity. The integrated photonics and circuit QED systems we consider satisfy the above requirements.

\begin{figure}[t!]
\centering
\subfigure[]{
{\includegraphics[scale=0.18]{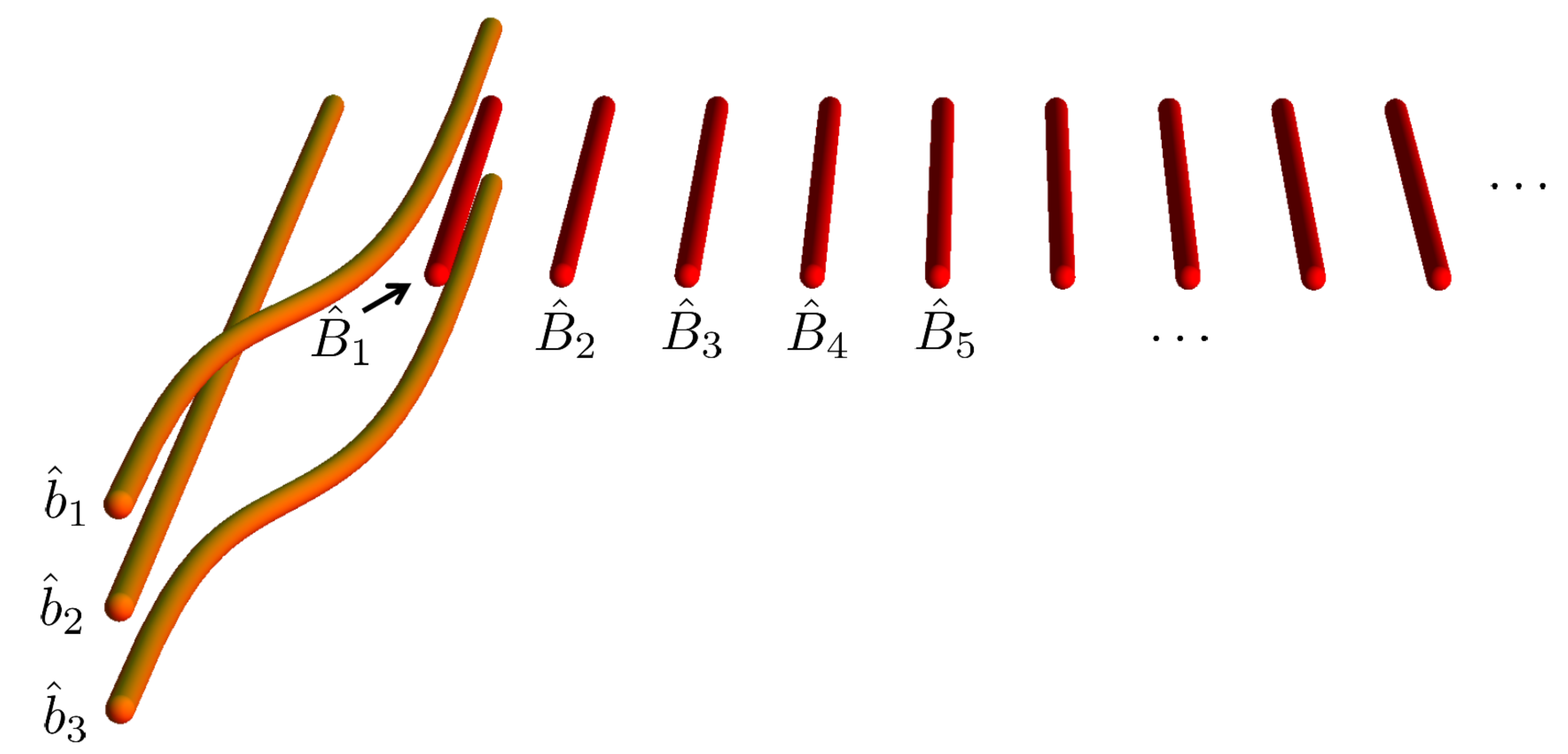}  \label{Agarwal_AfterJapan_N3}}
}
\subfigure[]{
{\includegraphics[scale=0.15]{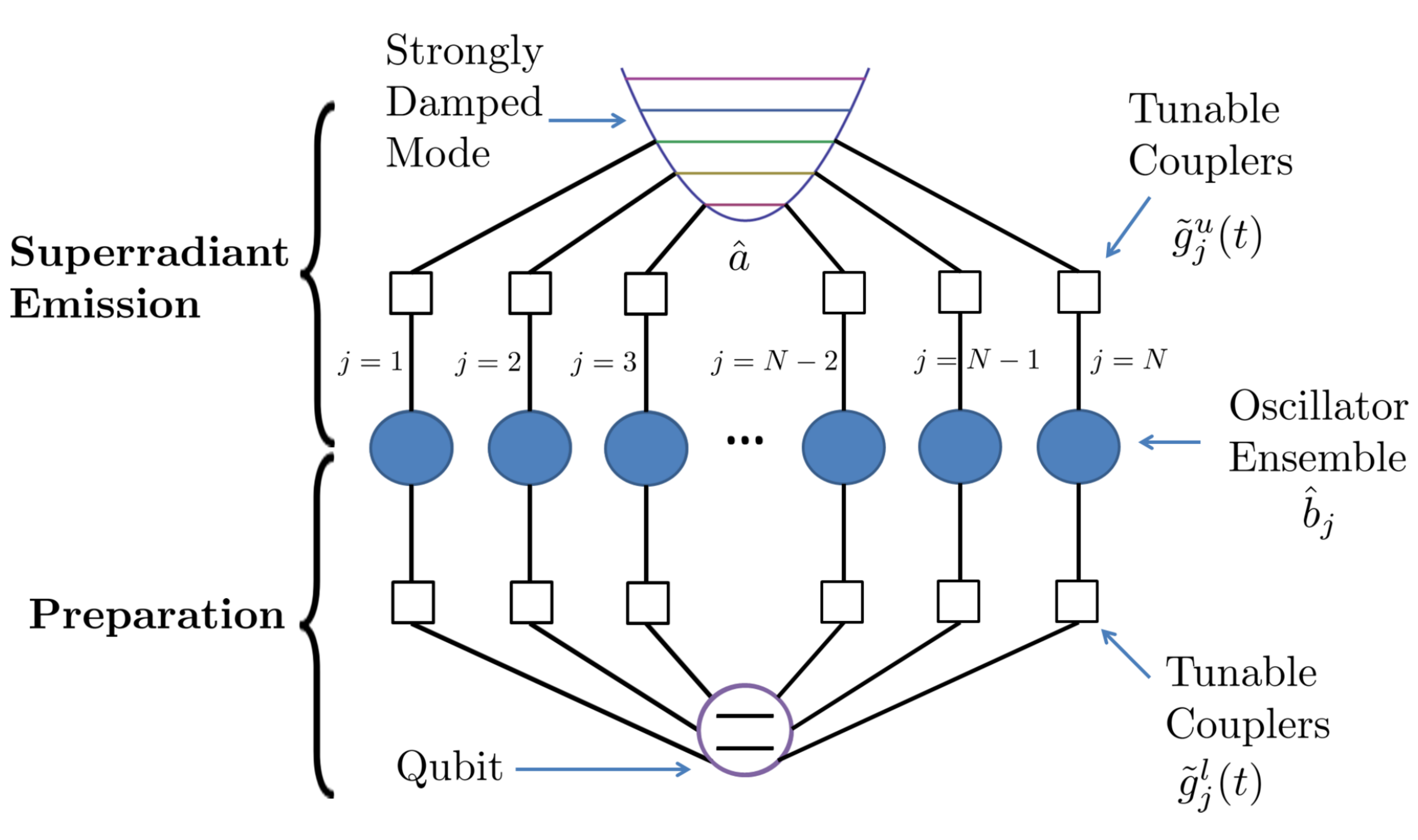} \label{circuitQEDDiagram}
}
}
\caption[]{ Implementations with initial state preparation a) Integrated Photonics b) Circuit QED. \textit{Integrated Photonics}: Three system modes (left) initially coupled to one another fan out and collectively couple to a linear array of bath waveguides (right). \textit{Circuit QED}:  strongly damped $\hat{a}$ mode for observation of superradiant emission (top) connected to tunable couplers (squares) which in turn connect to the system modes (circles).  The system modes are coupled to a qubit via another set of tunable couplers for initial state preparation.  
}

\end{figure}

\subsection{Integrated Photonics}

Here we consider an integrated photonics implementation as shown in Fig.~\ref{Agarwal_AfterJapan_N3}. The implementation consists of two sections; one for state preparation and the other for the observation of superradiance. The Hamiltonian for the state preparation section is,
\begin{eqnarray}\label{HamEvolutionWaveguideActual}
H/\hbar = \omega \sum^N_{j=1} \hat{b}^{\dagger}_j \hat{b}_j + J \sum^{N-1}_{j=1} ( \hat{b}^{\dagger}_j \hat{b}_{j+1} + \hat{b}^{\dagger}_{j+1} \hat{b}_j  ), 
\end{eqnarray}
where, $J$ is the coupling rate between system waveguides. For an initially unentangled input state, evolution under this Hamiltonian can entangle the modes. For example, for the input state $| \psi (0) \rangle = | n_1, n_2, \dots, n_N \rangle$ the evolved state is \cite{AgarwalWaveguidePaper},
\begin{eqnarray} \label{HamEvolutionWaveguide}
| \psi (t) \rangle = \prod^N_{j=1} \frac{1}{\sqrt{n_j !}} \left( \sum^N_{q=1} A_{j,q} b^{\dagger}_q    \right)^{n_j} |0\rangle^{\bigotimes N}
\end{eqnarray}
where 
\begin{eqnarray}
 \! \! \! \! \! \! \! \! A_{j,q}= \frac{2}{N+1} \sum^{N}_{p=1} e^{ \left( -i( \omega + 2J \cos (\frac{p \pi}{N+1})t )  \right)}  \sin(\frac{q p \pi}{N+1}) \sin(\frac{j p \pi}{N+1}). \nonumber
\end{eqnarray}

We now show how superradiant, normal and subradiant states can be generated in a three waveguide system by injecting a single photon into one of the modes. Using (\ref{HamEvolutionWaveguide}) we find that the states,  $| \psi_1 (0) \rangle = |1,0,0 \rangle$ and $| \psi_2 (0) \rangle = |0,1,0 \rangle$, will evolve into,
\begin{eqnarray}
| \psi_1 (t) \rangle &=& x \big( (z+1)^2  |1,0,0 \rangle - \sqrt{2} (z^2-1)  |0,1,0 \rangle \nonumber \\
&+& (z-1)^2  |0,0,1 \rangle \big), \\
| \psi_2 (t) \rangle &=& e^{-i\omega t} \big( -\frac{i}{\sqrt{2}} \sin (\sqrt{2}Jt) (|1,0,0 \rangle + |0,0,1 \rangle) \nonumber \\
&+& \cos (\sqrt{2}Jt) |0,1,0 \rangle \big),
\end{eqnarray}
where, $x= \frac{1}{4} e^{-i(\omega + \sqrt{2}J)t}$ and $z = e^{i \sqrt{2}J t}$. Assuming uniform coupling rates when these states enter the superradiant section of the device ($g_j = g \; \forall j$) the fraction of dark energy for these states are,
\begin{eqnarray}
F_1 &=& \frac{3}{4} - \frac{1}{12} \cos (2 \sqrt{2}J t) \ge F_N ,\\
F_2 &=& \frac{1}{2} + \frac{1}{6} \cos (2 \sqrt{2}J t)  \le F_N .
\end{eqnarray}
where, $ F_N= \frac{2}{3}$. For $t\neq n \pi /\sqrt{2} J, (n \in \mathbb{Z}^+)$ it is clear that injecting a single photon into the first waveguide results in a subradiant state and injecting a single photon into the second waveguide results in a superradiant state. Normal states can be prepared in either case for $t= n \pi /\sqrt{2} J$. At $t=t^*:= (n +\frac{1}{2}) \pi /\sqrt{2} J$, the first state is most subradiant ($F_1=5/6$) and the second state is most superradiant ($F_2=1/3$). Therefore we suggest engineering the length of the waveguides such that the Hamiltonian (\ref{HamEvolutionWaveguideActual}) evolves for $t=t^*$. We note that more complex states can be engineered by injecting more than one photon.

After the preparation section, the photons enter the second section for the observation of superradiance. The superradiance section consists of a semicircular array of system waveguides, $\hat{b}_j$, surrounding the end of a semi-infinite equispaced array of bath waveguides, $\hat{B}_k$. The Hamiltonian for this system is, 
\begin{eqnarray}\label{HIPwithBath}
\! \! \! \! \! \! \hat{H}/\hbar &=& \omega  \sum^N_{j=1} \hat{b}^{\dagger}_j \hat{b}_j  +  \mathcal{G}_N ( \hat{C}^{\dagger}_N \hat{B}_1 + \hat{B}^{\dagger}_1 \hat{C}_N  )\nonumber   \\
\! \! \! \! \! \! &+& \omega \sum^{\infty}_{k=1} \hat{B}^{\dagger}_k \hat{B}_k  + \frac{\kappa}{2} \sum^{\infty}_{k=1} ( \hat{B}^{\dagger}_k \hat{B}_{k+1} + \hat{B}^{\dagger}_{k+1} \hat{B}_k  )
\end{eqnarray}   
where, each waveguide has the same angular frequency $\omega$, the system waveguides couple to the first bath waveguide, $\hat{B}_1$, at the rates, $g_j$, and bath waveguides nearest-neighbor couple at the rate, $\kappa/2$. By design, dissipation within the waveguides is minimal and can be ignored on the timescale of interest \cite{OBrienIPReview,LonghiExperimentWithLosses, DirectWriteThorne}.

We make two fundamental assumptions; 1) that the system waveguides do not couple to one another and 2) that the bath waveguides couple to one another at a larger rate than the system modes couple to the bath, $\kappa \gg \mathcal{G}_N$. Physically the case of $N=3$ system waveguides is relatively easy to envisage (Fig.~\ref{Agarwal_AfterJapan_N3}), however,  additional system waveguides are problematic \footnote[8]{An opaque material could be placed between system waveguides whilst still allowing an interaction with the first bath waveguide.}. The second assumption requires that the distance between the system waveguides and the first bath waveguide is large compared to the distance between each bath waveguide. Starting from Hamiltonian  (\ref{HIPwithBath}), we can trace over the bath degrees of freedom to obtain the effective master equation of the system \cite{CarmichaelStatisticalMethods2}.  It can be shown that the effective master equation of the system is the superradiance master equation (\ref{rhocN}) \cite{chinesepaper,chinesePRA}. Experimentally the emission intensity, $I_N (t)$, is found as the sum of the intensity in each bath waveguide. Furthermore, correlation functions can also be measured in this system \cite{IPRandomWalks}.

\subsection{Circuit QED}

We now consider harmonic oscillator superradiance in a circuit QED system. We first introduce the system, then show how the dynamics are well described by the superradiance master equation and finally
describe how various initial states can be prepared in the system. The circuit QED system, depicted in Fig.~\ref{circuitQEDDiagram}, consists of two sections, the top section is for the observation of superradiance and the bottom section is for state preparation. The system is highly versatile as many different kinds of oscillators can be used, including stripline \cite{Blais2004} and lumped element resonators \cite{LumpedElement} and the transmon qubit operated in the harmonic parameter regime \cite{TransmonPaper}. The unitary dynamics of the circuit QED system is described by the Hamiltonian,
\begin{eqnarray}\label{cirQEDH}
\! \! \! \! \! \! \! \! \! \! \! \! \! \! \! \! \!  H/\hbar &=& \omega \hat{a}^{\dagger} \hat{a} +  \omega  \sum^N_{j=1} \hat{b}^{\dagger}_j \hat{b}_j  + \frac{\Omega (t)}{2} \hat{\sigma}_z + \mathcal{E}(t) \hat{\sigma}_+ + \mathcal{E}^*(t) \hat{\sigma}_- \nonumber \\
\! \! \! \! \! \! \! \! \! \! \!  \! \! \! \! \! \!  &+& \sum^N_{j=1} \tilde{g}^u_j (t) ( \hat{b}^{\dagger}_j  \hat{a} + \hat{a}^{\dagger} \hat{b}_j)
 + \sum^N_{j=1} \tilde{g}^l_j (t) ( \hat{b}^{\dagger}_j \hat{\sigma}_- + \hat{b}_j \hat{\sigma}_+ ),
\end{eqnarray} 
where from Fig.~\ref{circuitQEDDiagram},  the top oscillator represents the $\hat{a}$ mode, which couples to the $N$ central oscillators, $\hat{b}_j$, via the upper set of tunable couplers at the rates $\tilde{g}^u_j (t)$ \cite{tunablecouplerPRL,tunablecouplerPRA}. The central oscillators, $\hat{b}_j$, also couple to a superconducting qubit via lower set tunable couplers at the rates, $\tilde{g}^l_j (t)$. The qubit is driven at the complex amplitude, $\mathcal{E} (t)$, with tunable transition frequency, $\Omega (t)$. Dissipative dynamics are described by the master equation,
\begin{eqnarray} \label{MEcircuitQED}
\dot{\rho} &=& -i [H, \rho] + \frac{\kappa}{2}\mathcal{D}[a]\rho  +   \sum^N_{j=1}  \frac{\gamma_j}{2} \mathcal{D}[\hat{b}_j]\rho \nonumber \\ 
&+& \frac{\gamma^s_q}{2} \mathcal{D}[\hat{\sigma}_-]\rho + \frac{\gamma^p_q}{2} \mathcal{D}[\hat{\sigma}_z]\rho,
\end{eqnarray}
where, $\kappa$ is the decay rate of the strongly damped $\hat{a}$ mode and the collective modes  $\hat{b}_j$ decay at the rates, $\gamma_j$. Furthermore, the qubit undergoes energy relaxation at the rate $\gamma^s_q$ and dephases at $\gamma^p_q$. We assume that the dissipative rates, $\gamma_j, \gamma^s_q$, and  $\gamma^p_q$, are small compared to all other system parameters. This assumption is required to achieve high fidelity state preparation and in order to approximate the system dynamics by the superradiance master equation.

To observe superradiance it is necessary to prepare an initial state in the oscillators as discussed below, then decouple the qubit by either tuning the qubit's transition frequency far from $\omega$ or setting the lower tunable couplers to the off position, $\tilde{g}^l_j (t) = 0$. The system dynamics are therefore described by the master equation in (\ref{MEGTc}). Assuming the $\hat{a}$ oscillator is strongly damped, $\kappa \gg \mathcal{G}_N$, we can adiabatically eliminate the $\hat{a}$ degrees of freedom as in section \ref{SRSection} and arrive at the desired superradiance master equation (\ref{rhocN}).  Therefore harmonic oscillator superradiance can be observed by preparing the oscillator ensemble in a superradiant state and measuring the emission of photons into the $\hat{a}$ mode. 

We now describe a method to prepare highly entangled multimode states using the system in Fig.~\ref{circuitQEDDiagram}. For preparation, the $\hat{a}$ oscillator is decoupled by tuning the top row of couplers to their off state, $\tilde{g}^u_j (t)=0$. We consider a generalization of the Law and Eberly protocol to prepare entangled states of several modes \cite{LawandEberly}. The Law and Eberly protocol prepares an arbitrary single mode state (\ref{LandEstate}) by successively exciting the qubit into a superposition of the ground and excited states then partially transferring quanta from the qubit to the oscillator. The protocol requires the ability to switch between two interactions: 1) classical driving of the qubit for single qubit rotations and 2) a Jaynes Cummings interaction between the qubit and the mode. Furthermore, brief pauses are required between these two interactions to adjust the phase of the excited state of the qubit relative to the ground state \cite{MartinisLandE, PRL2Mode}. By varying either the coupling rates or the length of interaction time intervals, Law and Eberly demonstrated that an arbitrary single mode state (\ref{LandEstate}) could be prepared \cite{LawandEberly}.

We now discuss how each interaction can be generated in the circuit QED system in Fig.~\ref{circuitQEDDiagram}. The first interaction we require is the classical drive for qubit rotation. In the interaction picture, qubit rotations are generated by evolution under the Hamiltonian, \begin{equation}
\hat{H}_d = \mathcal{E}\hat{\sigma}_+ + \mathcal{E}^* \hat{\sigma}_-,
\end{equation}
where the drive is set to a constant amplitude, $\mathcal{E}(t) = \mathcal{E}$.
This Hamiltonian can be obtained from (\ref{cirQEDH}) by two different methods. Either the tunable couplers could be switched off ($\tilde{g}^l_j (t) = 0$) or the qubit transition frequency could be tuned far out of resonance, $\Omega (t)\gg \omega$ or $\Omega (t)\ll \omega$ \cite{MartinisLandE}.

The second interaction we require is the Jaynes Cummings interaction. In order to attain the Jaynes Cummings interaction, the qubit is tuned into resonance with the modes, i.e $\Omega (t) = \omega$, and the tunable couplers are tuned to a predefined constant rate $\tilde{g}^l_j (t)=\tilde{g}_j$. In the interaction picture the Hamiltonian (\ref{cirQEDH}) now describes the Jaynes Cummings interaction,
\begin{equation}
\hat{H}_{JC} =  \tilde{g} ( \hat{d}^{\dagger} \hat{\sigma}_- + \hat{d} \hat{\sigma}_+ ),
\end{equation}
where, $\hat{d} = \sum^N_{j=1} \tilde{g}_j \hat{b}_j/ \tilde{g}$, $\tilde{g}= \sqrt{\sum^N_{j=1} \tilde{g}^2_j}$ and $[\hat{d},  \hat{d}^{\dagger}]=1$. We note that $\hat{d}$ can represent a variety of modes, including $\hat{C}_j$, where $j=1,2,\dots, N$. The use of the collective mode, $\hat{d}$, rather than a single mode, $\hat{b}_j$, is key to our proposal. 

Using the Law and Eberly protocol, sequentially alternating between evolutions under $\hat{H}_d$ and $\hat{H}_{JC}$, with brief pauses to correct the phase between the ground and excited state of the qubit, it is possible to prepare the state,
\begin{equation}
| \psi \rangle  = \sum^{N_1}_{n=0} c_n  \frac{(\hat{d}^{\dagger})^n}{\sqrt{n!}} | 0 \rangle^{\bigotimes N}, 
\end{equation}
where, $c_n$ are arbitrary complex amplitudes with $\sum^{N_1}_{n=0} |c_n|^2 =1$ and $N_1$ is the maximum Fock state, which is limited due to dissipation. For example, when $\hat{d} = \hat{C}_N$, the state, $ \sum^{N_1}_{R=0} c_R | \mathbf{e}_0,  \Phi^R_0 \rangle $, can be prepared. For large $N_1$ and small $|\xi|$ this state could approximate the collective squeezed vacuum $|  \Phi^{\xi}_0 \rangle$. Similarly,  when $\hat{d} = \hat{C}_k \; (k\neq N)$, we can prepare the state, $ \sum^{N_1}_{L=0} c_L | L \mathbf{e}_k,  \Phi^0_k \rangle $. Furthermore, for $\hat{d} = (\hat{C}_k + \hat{C}_N)/\sqrt{2}, (k\neq N)$ and $c_n = \delta_{n,2}$ the state,
\begin{eqnarray}
\! \! \! \! \! \! \! \! \frac{(\hat{d}^{\dagger})^2}{\sqrt{2}} | 0 \rangle^{\bigotimes N}
 &=& \frac{1}{2\sqrt{2}} \left(  \sqrt{2} | \mathbf{e}_0, \Phi^2_0 \rangle +  \sqrt{2} | 2 \mathbf{e}_k, \Phi^0_2 \rangle +  2| \mathbf{e}_k, \Phi^1_1 \rangle  \right) \nonumber
\end{eqnarray}
can be prepared.

We have shown that a variety of multimode entangled states can be prepared using the circuit QED system in Fig.~\ref{circuitQEDDiagram}. These states are useful in the study of harmonic oscillator superradiance as they include several of the collective states studied in section \ref{SRIc}. In addition, multimode states prepared by the generalized Law and Eberly protocol could be useful for metrology \cite{MetrologyPaper}.

\section{Conclusion \label{conclusion}}

In summary, we have studied superradiance in an ensemble of harmonic oscillators. We have shown that harmonic oscillator superradiance should be observable in a variety of systems including waveguide arrays and superconducting resonators. The bosonic Dicke basis was introduced which proved very convenient in calculations, particularly in classifying dark states and in analyzing the dynamics of the system.  We also defined a superradiance criterion, which can be used to classify arbitrary harmonic oscillator ensemble states as superradiant, normal or subradiant.

Furthermore we characterized a range of states including bosonic Dicke basis states, multimode Fock states, incoherent mixtures, thermal states, single mode squeezed coherent states, collective coherent states and collective squeezed coherent states. For each state the emission intensity, fraction of dark energy and two-time correlation functions were calculated. One of the most interesting findings was that single mode squeezing has normal radiance, whereas squeezing in the $N$-th collective mode was superradiant. Finally we generalized the Law and Eberly protocol to create highly entangled multimode states in circuit QED. The protocol should allow the preparation of many of the states discussed.

We anticipate our results to be useful in further studies of harmonic oscillator superradiance and other  multimode phenomena. The systems we considered may have applications in continuous variable quantum computing and the decoherence free subspaces may be useful as quantum memories. In addition, the state preparation protocol may produce states useful for metrology. Future work will consider entanglement and the effect of dissipation on the dynamics in this system.

\begin{acknowledgments}
We thank L. Willink and M. S. Kim for useful discussions.  This work was supported by the European Commission projects QUANTIP \# 244026 and Q-ESSENCE \#248095. S. R. would like to thank ARC (DP1094758) for financial support.
\end{acknowledgments}

\appendix

\section{Eigenstates}
In this section we diagonalize the Hamiltonian (\ref{HGTc}) in the bosonic Dicke basis introduced in section \ref{system}. In order to find the eigenstates we split the Hamiltonian into two commuting parts, 
\begin{equation} \label{TotHamApp}
\hat{H} =  \omega \hat{L} + \hat{H}'
\end{equation}
where,
\begin{equation}
 \hat{H}' =  \omega \hat{a}^{\dagger} \hat{a} + \omega \hat{R}+ \mathcal{G}_N (\hat{a}^{\dagger} \hat{C}_N + \hat{a} \hat{C}_N^{\dagger}  ).
\end{equation}

We now introduce the normal modes,
\begin{eqnarray}
\hat{d}_{\pm} &=& \frac{1}{\sqrt{2}}(\hat{a} \pm \hat{C}_N) 
\end{eqnarray}
where, $[\hat{d}_i, \hat{d}^{\dagger}_j]=\delta_{i,j}$. In terms of these normal modes $\hat{H}'$ is diagonal,
\begin{equation}
 \hat{H}' =  (\omega+ \mathcal{G}_N ) \hat{d}_+^{\dagger} \hat{d}_+ + (\omega- \mathcal{G}_N ) \hat{d}_-^{\dagger} \hat{d}_-
\end{equation}

Therefore the eigenstates of the Hamiltonian (\ref{TotHamApp}) are simultaneous eigenkets of $\hat{L}$ and $\hat{H}'$,
 \begin{eqnarray}
\hat{H} | l, n_+, n_- \rangle  &=& E^l_{n_+,n_-} | l, n_+, n_- \rangle 
\end{eqnarray}
where,  $E^l_{n_+,n_-}= \omega (l+n_+ + n_- ) +  \mathcal{G}_N  (n_+- n_-)$ and $n_{\pm} = \langle \hat{d}^{\dagger}_{\pm} \hat{d}_{\pm} \rangle $.
The eigenstates are explicitly,
\begin{equation}
| l, n_+, n_- \rangle  = \frac{( \hat{d}^{\dagger}_+)^{n_+} ( \hat{d}^{\dagger}_-)^{n_-}}{\sqrt{n_+! n_- !}} |0, \Phi^0_{l} \rangle.
 \end{equation}

\section{Density Matrix Elements of the Superradiance Master Equation for Pure Initial States \label{MatrixElementsSRMEAppendix}}

We now solve for the matrix elements and emission intensity of (\ref{rhocN}) given an arbitrary pure initial state of the $\hat{b}_j$ modes, 
\begin{equation}\label{Ic2}
| \psi (0) \rangle =  \sum^{\infty}_{L,R=0} \sum_{\mathbf{d}_L} a_{(R, L,\mathbf{d}_L)}  |\mathbf{d}_L, \Phi^R_L \rangle,
\end{equation}
where, $\sum^{\infty}_{L,R=0} \sum_{\mathbf{d}_L} |a_{(R, L,\mathbf{d}_L)}  |^2 =1$. To proceed we introduce the matrix elements, $P_{(R, L,\mathbf{d}_L)} (t)= \langle \mathbf{d}_L, \Phi^R_L | \rho_b (t) | \mathbf{d}_L, \Phi^R_L   \rangle $. In terms of these matrix elements the superradiance master equation is transformed into the difference differential equation,
\begin{equation}\label{DifferenceEqnNew}
\frac{d }{d t} P_{(R, L,\mathbf{d}_L)} (t) = N \Gamma \left( (R+1)P_{(R+1, L,\mathbf{d}_L)} (t) -  R P_{(R, L,\mathbf{d}_L)} (t)  \right).
\end{equation}

We can re-express the difference equation (\ref{DifferenceEqnNew}) using vector notation, 
\begin{equation}\label{MatEqn}
\dot{\mathbf{P}}(t) = N \Gamma A \mathbf{P}(t),
\end{equation}
where, $\mathbf{P}(t) = (  P_{(0, L,\mathbf{d}_L)} (t),   P_{(1, L,\mathbf{d}_L)} (t),   P_{(2, L,\mathbf{d}_L)} (t), \dots )^{T}$, $\mathbf{P}_n (t) = P_{(n-1, L,\mathbf{d}_L)} (t)$, and,
 \[
A_{i,j} = \left\{ 
  \begin{array}{l l}
(1-i) & \quad \text{if $i=j$}\\
(j-1) & \quad \text{if $j=i+1$}\\
    0 & \quad \text{otherwise.}\\
  \end{array} \right.
\]
To solve the system of ODE's (\ref{MatEqn}) we diagonalize $A$ using the Pascal matrix \cite{pascal}, 
\[
  B_{i,j} = \left\{ 
  \begin{array}{l l}
    \binom{i-1}{j-1} & \quad \text{if $i\ge j\ge1$}\\
    0& \quad \text{ otherwise.}\\
  \end{array} \right.
\]
After some manipulation it can be shown that, $A = (B^{-1})^T D B^T$ where, $D_{i,j} = \delta_{i,j} (1-i)$. Therefore the eigenvalues of $A$ are $\lambda_i = (1-i)$ and the eigenvectors, $v_i$, are the columns of $ (B^{-1})^T$. Hence for a given vector of initial conditions $\mathbf{P}^0 = ( |a_{(0, L,\mathbf{d}_L)}|^2, |a_{(1, L,\mathbf{d}_L)}|^2, \dots  )^T$, where $\mathbf{P}^0_k =  |a_{(k-1, L,\mathbf{d}_L)}|^2$ , the solution to (\ref{MatEqn}) is,
\begin{equation}
 \mathbf{P}(t) = \sum^{\infty}_{i=1} c_i e^{\lambda_i  N \Gamma t} v_i, 
\end{equation}
where, the coefficients $\mathbf{c}=(c_1, c_2, \dots)^T$ can be determined from the initial condition,
\begin{equation}
 \mathbf{c} = B^T \mathbf{P}^0.
\end{equation}

We can rewrite the $n$-th element of this solution as,
\begin{eqnarray}
\! \! \! \! \! \! \mathbf{P}_n (t)  &=&  \sum^{\infty}_{i=n}  \sum^{\infty}_{k=1} \mathbf{P}^0_k  e^{\lambda_i  N \Gamma t} (-1)^{i-n}   \binom{i-1}{n-1} \binom{k-1}{i-1} \nonumber \\
\! \! \! \! \! \!  &=&  \sum^{\infty}_{k=n}  \mathbf{P}^0_k e^{(1-n)  N \Gamma t} (1-e^{- N \Gamma t})^{k-n}  \binom{k-1}{n-1}
\label{ProbsOneLadder}
 \end{eqnarray}
where, $ \mathbf{P}_n (t) = P_{(n-1, L,\mathbf{d}_L)} (t)$ and $\mathbf{P}^0_k = ( |a_{(k-1, L,\mathbf{d}_L)}|^2$. This result can also be written in terms of the matrix elements (instead of using vector notation),
\begin{equation} \label{imp1}
P_{(R, L,\mathbf{d}_L)} (t) = \sum^{\infty}_{k=R}  | a_{(k, L,\mathbf{d}_L)} |^2 e^{-R  N \Gamma t} (1-e^{-  N \Gamma t})^{k-R}  \binom{k}{R}.
\end{equation}

For example for the single bosonic Dicke basis state, $| \psi (0)\rangle = | \mathbf{d}_L, \Phi^K_L \rangle$, 
\begin{equation}  \label{PopSingleBosonicDicke}
P_{(R, L,\mathbf{d}_L)} (t) = \binom{K}{R} e^{-R  N \Gamma t} (1-e^{-  N \Gamma t})^{K-R}.
\end{equation}
The populations for five oscillators prepared in $| \mathbf{e}_0, \Phi^5_0 \rangle$ are displayed in Fig.~\ref{POPDiagramF}.

\section{Superradiance of Atoms}

In order to compare the superradiance of oscillators to the superradiance of atoms, we briefly review some relevant results \cite{MySRpaper, Lee1}. Consider a system of $N$ two-level atoms coupled to a strongly damped mode, $\hat{a}$. In the interaction picture the system evolves under the Tavis-Cummings Hamiltonian,
\begin{equation} \label{HNcPBcollect}
H = g ( \hat{J}_- \hat{a}^{\dagger} + \hat{a} \hat{J}_+),
\end{equation}
where, all atoms couple to the field at the rate, $g$, and the collective operators are $\hat{J}_{\pm} = \sum^N_{i=1} \hat{\sigma}^{\pm}_i$. The master equation for this system is,
\begin{eqnarray} \label{ME_NcPB}
 \dot{\rho} &=& -i [ \hat{H}, \rho] +\frac{\kappa}{2}  \mathcal{D}[\hat{a}] \rho + \sum^N_{j=1} \frac{\gamma^s_j}{2} \mathcal{D}[ \hat{\sigma}_j^-] \rho + \frac{\gamma^p_j}{2}  \mathcal{D} [\hat{\sigma}_j^z]  \rho \nonumber
\end{eqnarray}
where $\gamma^s_j$ and $ \gamma^p_j$ are for the $j$th atom the spontaneous emission and dephasing rate respectively and $\kappa$ is the field decay rate. Similar to section \ref{SRSection}, we assume $\kappa \gg g \gg \gamma^s_j, \gamma^p_j $ and adiabatically eliminate the field degrees of freedom to arrive at the superradiance master equation \cite{MySRpaper, SR_PRLWoggon},
\begin{eqnarray} 
\dot{\rho}_q =   \frac{ 2 g^2 }{\kappa}  \mathcal{D}[\hat{J}_-] \rho_q.
\label{AtomSRME}
\end{eqnarray}
Expanding the superradiance master equation in the atomic Dicke basis, we obtain the following equation for the matrix elements,
\begin{eqnarray}\label{DifferenceEqn}
\dot{P}(l,m,t) &=& \Gamma [ (l-m)(l+m+1) P(l,m+1,t) \nonumber \\
&-& (l+m)(l-m+1) P(l,m,t)  ].
\end{eqnarray}
where, $P(l,m,t) =  \langle l,m | \rho_q (t) |l,m \rangle$ and $\Gamma =\frac{ 4 g^2 }{\kappa}$ is the single atom decay rate. Assuming the atoms are initially all excited, $P(l,m,0)=P(N/2,N/2, 0)$, and noting $l$ is conserved by the superradiance master equation (\ref{AtomSRME}), we can make a change of variable to $n=l-m= 0, 1, ..., N$. Here, $n$, corresponds to the number of photons emitted from the ensemble. In terms of $n$ (\ref{DifferenceEqn}) becomes,
\begin{eqnarray}\label{DifferenceEqn2}
\dot{P}(n,\tau) &=&  (N-n+1) n P(n-1,\tau) - (N-n)(n+1) P(n,\tau),\nonumber
\end{eqnarray}
where, we have rescaled the time $\tau =\Gamma t$. Using the Laplace transform the matrix elements are found to be,
\begin{equation}
\label{Lap2}
P(n,\tau ) = \mathcal{L}^{-1} \{ \frac{1}{  s+N} \prod^{n}_{i=1} \frac{(N-i+1)i}{ s+ (N-i)(i+1)} \}.
\end{equation}
where, $ \mathcal{L}^{-1} \{ \hat{f} (s) \} = f (t)$, is the inverse Laplace transform. For example for $N=5$ the populations are,
\begin{subequations}
\label{DickeStatePops}
\begin{eqnarray}
P(0, \tau) &=& e^{-5 \tau}  \\
P(1, \tau) &=& -\frac{5}{3} e^{-8 \tau}+\frac{5 e^{-5  \tau}}{3} \\
P(2, \tau) &=& -\frac{40 e^{-8   \tau}}{3} + 10 e^{-9   \tau}+\frac{10 e^{-5   \tau}}{3} \\
P(3, \tau) &=& -90 e^{-9   \tau}+80 e^{-8   \tau}+10 e^{-5   \tau} \nonumber \\
&-& 120 e^{-8   \tau}   \tau \\
P(4, \tau) &=&  -\frac{220 e^{-5   \tau}}{3}+ 180 e^{-9   \tau}-\frac{320 e^{-8   \tau}}{3} \nonumber \\
&+& 320 e^{-8   \tau}   \tau+80 e^{-5  \tau}  \tau \\
P(5, \tau) &=& 1-100 e^{-9   \tau}+\frac{125 e^{-8   \tau}}{3}\nonumber \\
&+& \frac{172 e^{-5   \tau}}{3} - 200 e^{-8   \tau}   \tau-80 e^{-5  \tau}   \tau. 
\end{eqnarray}
\end{subequations}
These populations are plotted in Fig.~\ref{POPDiagramA}, where we use the substitution, $n=5-K$. The emission intensity can be found from the populations via $I_N(\tau) = \frac{\partial}{\partial \tau} \sum^N_{n=0} n P(n,\tau)$. For example for $N=5$ the emission intensity is,
\begin{equation} \label{IntensitySingleAtomicDickeState5}
I_5 (\tau) = \frac{5}{3} [162e^{-9 \tau} +16 e^{-8 \tau} (24 \tau -1)  + e^{-5 \tau} (240 \tau-143)],
\end{equation}
which is shown in Fig.~\ref{FiveLevelDecay}.

%\bibliography{collectivePRABIB}

\end{document}